\documentclass[epj,nopacs]{svjour}
\usepackage[utf8]{inputenc}
\usepackage{amsmath,amssymb,amsfonts,graphicx,cite,multirow}

\newcommand{\Herwig}{{Herwig}}
\newcommand{\Pythia}{{Pythia}}
\newcommand{\PROFESSOR}{{Professor}}
\newcommand{\ptmin}{p_{\perp}^{\rm min}}
\newcommand{\dd}{\mathrm{d}}

\usepackage{color}
\usepackage{hyperref}
\usepackage{dsfont} 
\usepackage{booktabs,tabularx,ragged2e,caption,rotating}

\newcommand{\avg}[1]{\left \langle #1 \right \rangle }

\preprint{KA-TP-42-2016\\MCNET-16-46}

\title{Soft and diffractive scattering with the cluster model in Herwig}

\author{Stefan Gieseke, Frash\"{e}r Loshaj, Patrick Kirchgae\ss{}er}

\institute{Institute for Theoretical Physics, Karlsruhe Institute of
  Technology, Wolfgang-Gaede-Str.~1, 76131 Karlsruhe, Germany.}

\date{December 14, 2016}

\abstract{We present a new model for soft interactions in the
  event-generator \Herwig{}.  The model consists of two components.
  One to model diffractive final states on the basis of the cluster
  hadronization model and a second component that addresses soft
  multiple interactions as multiple particle production in
  multiperipheral kinematics.  We present much improved results for
  minimum-bias measurements at various LHC energies.  
  \PACS{{xx.yy.zz}{Xx Yy Zz}} }

\begin{document}
\maketitle
\section{Introduction}

With increasingly precise data on observables related to jet physics at
the LHC, the impact of soft physics on the accurate modeling of final
states plays an increasingly important r\^{o}le.  While on the one hand
there is enormous progress on the perturbative side, soft physics is far
from a systematically improvable description and Monte Carlo event
generators \cite{Bahr:2008pv,Sjostrand:2007gs,Gleisberg:2008ta} merely
resort to pure modeling of final states.  There is, however, a vast
theoretical and phenomenological knowledge on soft physics, accumulated
over the last decades that should at least be the benchmark for the
modeling in modern event generators.  

Still, a large effort is made to model soft aspects of event generation
at the LHC \cite{Fischer:2016zzs,Rasmussen:2015qgr,Christiansen:2015yca,Bierlich:2015rha,Hoche:2007hg}.
With the rise of increasingly accurate data, these aspects become
important again as with the increasing precision on the perturbative
side of event simulation, non-perturbative aspects become an important
part of the uncertainties.

In addition to the interest in modeling collider physics accurately,
soft phyics is interesting in its own right.  Furthermore, it plays an
important r\^{o}le for the understanding of cosmic ray data.

\subsection{The MPI model in Herwig}

Minimum-bias interactions at hadron colliders have always been on the
edge of physical modeling in \Herwig{}.  While there has been a model
for the underlying event (UE) that has been an add-on to the Fortran version
of the program, the so called \textsc{jimmy} package
\cite{Sjostrand:1987su,Butterworth:1996zw}, a similar model, based on
multiple partonic interactions (MPI) had been integrated into the newer
C++ version of \Herwig{} later \cite{Bahr:2008wk,Bahr:2008dy}.  In
addition to the hard multiple interaction model, there have also been
soft interactions included in the so-called hot-spot model
\cite{Borozan:2002fk,Bahr:2009ek}.

Here, semi-hard multiple interactions are included as multiple partonic
interactions into the existing framework of partonic scattering with
full parton showers and hadroni\-zation as for the initial hard process.
The interactions are based on the same assumptions as partonic
interactions in the usual collinear factorization approach, where for
the UE they are modeled in \Herwig{} with parton
distribution functions that have the valence quark contribution
subtracted.  The idea is that the main triggering process usually
terminates its parton shower with the extraction of a valence quark and
hence the probability to extract yet another valence quark is
suppressed.  In contrast, the backward evolution of the initial state
partons in secondary interactions will always end on a gluon, extracted
from the projectile hadron.  In this way, the colour structure of the
secondary scatters is unambiguous, which in turn is mandatory to hook up
a hadronization model.  There are two important parameters of this model.
$\ptmin$ is the transverse momentum at which the differential scatting
spectrum is cut-off in the infrared and $\mu^2$, which characterizes the
inverse proton radius for the transverse spatial distribution of partons
within the hadron, which in turn is taken from the dipole form factor.

The additional soft scatters are modeled as another type of multiple
interaction with a spatial distribution of the same functional form but
another, independent, inverse radius $\mu^2_{\rm soft}$ for soft
particles, which are usually broader than the hard partons.  Furthermore,
the transverse momentum spectrum is modeled with a Gaussian below
$\ptmin$,
\begin{equation}  
  \frac{\dd \sigma}{\dd p_{\perp}} = 
  \left(\frac{\dd \sigma_{\rm hard}}{\dd p_{\perp}}\right)_{p_{\perp}=\ptmin}
  \left(\frac{p_{\perp}}{\ptmin}\right) e^{-\beta\left( p_{\perp}^2 -\ptmin{}^2\right)}\ . 
      \label{eq:gaussian}
\end{equation}
Here, the spectrum is chosen such that it is continuous at $\ptmin$.  The
two parameters $\mu^2_{\rm soft}$ and $\beta$ are fixed from the
additional constraints that the total cross section 
\begin{equation}
  \label{eq:sumtot}
  \sigma_{\rm tot} = \sigma_{\rm hard}^{\rm inc} + \sigma_{\rm soft}^{\rm inc}
\end{equation}
and the elastic slope parameters are given by their known values, of
which good parametrizations are available.  

With this model for hard and soft MPI in \Herwig{} it has always been
possible to give a good description of the UE, i.e.\ the
presences of multiple interactions within one triggered hard
scattering.   In addition, also minimum-bias interactions have been
modeled.  Here, due to the setup of \Herwig{} that every event is being
triggered by one hard interaction, a dummy process is set up, where two
quarks with zero transverse momentum are pulled out of the proton, such
that then secondary hard and soft scatters give rise to a description of
a minimum-bias event.   This model gives satisfactory results
whenever the hard contributions dominate.  Measurements of fiducial
cross sections at the LHC here then cut either on low transverse
momentum particles or require a minimum number of charged particles in
order to suppress contributions from typical diffractive final state
signatures.  

\subsection{Breakdown of the MPI model in Herwig}

When applied to fiducial measurements, where these cuts are loosened,
the description of minimum-bias events with \Herwig{} is bound to fail.
The reason is that in particular the model for soft interactions is very
much ad-hoc.  It will give the production of soft particles in a way
that the 'turn-on regions' in the UE measurements are
well-described, but not the correlations among them or with other
hard particles.  So, the soft model is limited to describe the average
soft activity that accompanies a hard event.  

This failure is clearly visible, when our model for mini\-mum-bias events
is applied to observables which have prominent contributions from
diffractive events, or are even designed to emphasize contributions of
these.  This can of course be done, albeit there has never been a claim
from the \Herwig{} authors that these measurements could be described,
as, clearly, there has so far not been a model for diffractive events.

The prominent example is the so-called 'bump' problem, which was first
observed by ATLAS \cite{Aad:2012pw}.  The measurement finds the
distribution of large gaps in pseudorapidity $\Delta \eta_F$ in the
forward region of the detector in events with a minimal trigger.
$\Delta\eta_F$ is the larger of the pseudorapidity gaps from either end of the
tracker to the track with the largest (smallest, resp.) pseudorapidity.
\Herwig{} is found to over-emphasize the region of large gaps which is
mostly attributed to diffractive event topologies.  A closer inspection
has shown that these events stem from high-mass clusters that stretch
out into the forward regions and can be attributed to the colour
assignment of non-perturbative partons that are produced in the decay of
the proton remnants \cite{Gieseke:2016pbi}.

\subsection{New model for soft interactions}

The bump problem together with other shortcomings of the simulation of
relatively soft particle production lead us to rethink the model of
soft interactions in \Herwig{}.  There is on the one hand the lack of
simulation of diffractive final states and on the other hand the model
for soft interactions which seems to be very ad-hoc.  More hints for
problems with soft interactions can be seen in the soft part of
transverse momentum spectra of charged particles or identified hadrons
which show a pronounced structure of a suppression of soft particles in
the region of $p_{\perp}\sim 1\,$GeV. 

In this paper we introduce a model for diffractive final states, based
on the cluster hadronization model.  The idea is to make use of the
phenomenological parametrization of diffractive cross sections in a
Gribov--Regge factorization approach in order to produce diffractive
systems with certain momentum transfer $t$ and diffractive mass $M$ and
couple these with the cluster hadronization. 

The second new model concerns the production of soft particles.  From
observations we expect soft particle production to be connected with a
particle production which is flat in rapidity and quite narrow in
transverse momentum.  These requirements are fulfilled by the usual
models for soft gluon production, based on small-$x$ dynamics
\cite{Ciafaloni:1987ur,Catani:1989sg,Catani:1989yc,Marchesini:1994wr,Jung:2001hx,Balitsky:1978ic}.

In the following Sec.~\ref{sec:diffraction} we introduce the new
diffraction model, followed by the model for soft particle production in
Sec.~\ref{sec:soft}.  We tune the parameters of these models in
conjunction with other sensitive parameters of the remaining parts of
the MPI model and describe this procedure in Sec.~\ref{sec:tuning},
before we present first results in Sec.~\ref{sec:results}.

\section{Diffraction model}
\label{sec:diffraction}
In this section we describe in more detail the implementation of high-mass diffraction dissociation within the cluster model which was initially presented
in \cite{Gieseke:2016pbi}. Events are generated utilizing differential
cross sections for single and double diffraction only, where central
diffraction remains to be implemented in the future. These cross
sections can be derived from Regge theory and the generalized optical
theorem, or the so-called Mueller's theorem \cite{Mueller:1970fa} (for a
review see for example \cite{Barone:2002cv}). Let's consider the single
diffractive dissociation process $A+B\rightarrow X+B$ first,
where $A$ and $B$ are hadrons and $X$ is some hadronic final state, in
the limit $s\gg M^2\gg |t|$. $s$ is the total center of mass energy of the
incoming particles, $M$ is the invariant mass of the state $X$ and $-t$
is the momentum transfer. In this work we focus only in the case where
both hadrons $A$ and $B$ are protons. By considering the amplitude for a
single Pomeron exchange, linearity of Regge trajectory and generalized
optical theorem, we can write for single diffraction:
\begin{align}
\frac{d^2\sigma^{SD}}{dM^2dt} =& \frac{g_{3\mathds{P}}(0)}{16\pi^2
                                 s}|g_\mathds{P}(t)|^2 g_\mathds{P}(0)
                                 \nonumber \\
                               &\times\left(\frac{s}{M^2}\right)^{2\alpha_\mathds{P}(t)-1}\left(M^2\right)^{\alpha_\mathds{P}(0)-1},
\label{eq:sdcs}
\end{align}
where $g_\mathds{P}$ and $g_{3\mathds{P}}$ are the proton-pomeron and
the triple pomeron coupling respectively and they are in general $t$ dependent. For small values of $|t|$, 
\begin{align}
\frac{d^2\sigma^{SD}}{dM^2dt} = N \left(\frac{s}{M^2}\right)^{\alpha_\mathds{P}(0)} e^{\left( B_0 + 2 \alpha' \ln\frac{s}{M^2}\right)t},
\label{eq:sdcs1}
\end{align}
where $B_0 \approx 10.1\,\mathrm{GeV}^{-2}$ is the so-called proton-pomeron
slope; the normalization constant depends on the proton-pomeron and
triple pomeron coupling. We have also used the linearity of the Regge
trajectory, $\alpha_{\mathds{P}}(t)=\alpha_{\mathds{P}}(0)+\alpha' t$,
where $\alpha'$ and $\alpha(0)$ are the pomeron slope and intercept
respectively.

For double diffraction $A+B\rightarrow X_A + X_B$, one can derive the
differential cross section using the factorization property of Regge
amplitudes and then use the results for single diffraction and elastic
cross sections. For details see references mentioned above. The result
is:
\begin{align}
\frac{d^3\sigma^{DD}}{dM_A^2dM_B^2dt} = &\frac{1}{16\pi^3 s} g_\mathds{P}^2(0) g_{3\mathds{P}}^2(0) \left(\frac{s}{M_A^2M_B^2}\right)^{2\alpha_\mathds{P}(t)-1}\nonumber \\ &\times\left(M_A^2\right)^{\alpha_\mathds{P}(0)-1}\left(M_B^2\right)^{\alpha_\mathds{P}(0)-1}.
\label{eq:ddcs1}
\end{align}  
Similarly as above, for small momentum exchange one can write:
\begin{align}
\frac{d^3\sigma^{DD}}{dM_A^2dM_B^2dt} =& N
                                         \left(\frac{s}{M_A^2}\right)^{\alpha_\mathds{P}(0)}\left(\frac{s_0}{M_B^2}\right)^{\alpha_\mathds{P}(0)}
  \nonumber \\
  &\times e^{\left( b + 2 \alpha' \ln\frac{s s_0}{M_A^2M_B^2}\right)t},
\label{eq:ddcs1}
\end{align}
where $s_0$ is fixed in the total normalization and $b$ is a constant
set to $\sim 0.1$.  The total and relative normalization between single
and double diffraction is not fixed and it is chosen roughly according
to measurements of total cross sections in \cite{Abelev:2012sea}.

In order to integrate the diffractive model into the MPI model in
\Herwig{}, we have to ensure that the cross sections for hard and soft
interactions only sum up to a fraction of the total cross section when
we fix the model parameters of the soft interaction
\cite{Bahr:2009ek,Bahr:2008pv}.  We assume that the diffractive events
come at a rate of about $20-25\%$ of the total event rate.  Then, we can
generate the diffractive processes as an independent sample.

\begin{figure}[t]
\centering
\includegraphics[width=5cm]{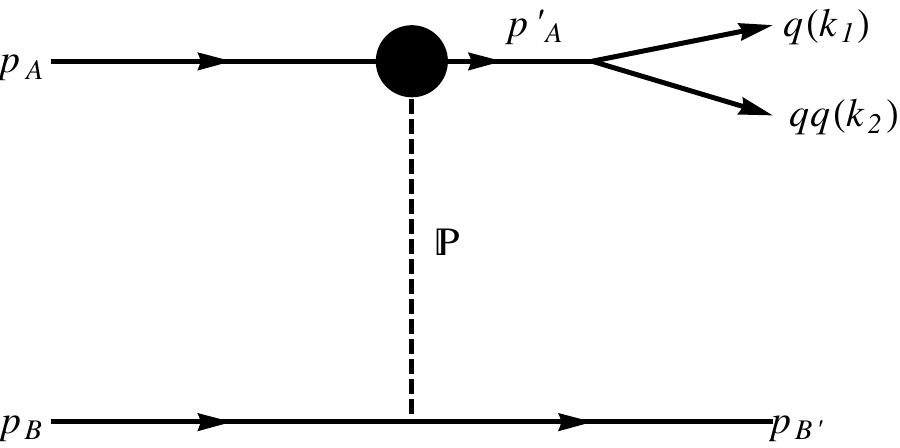}
\includegraphics[width=5cm]{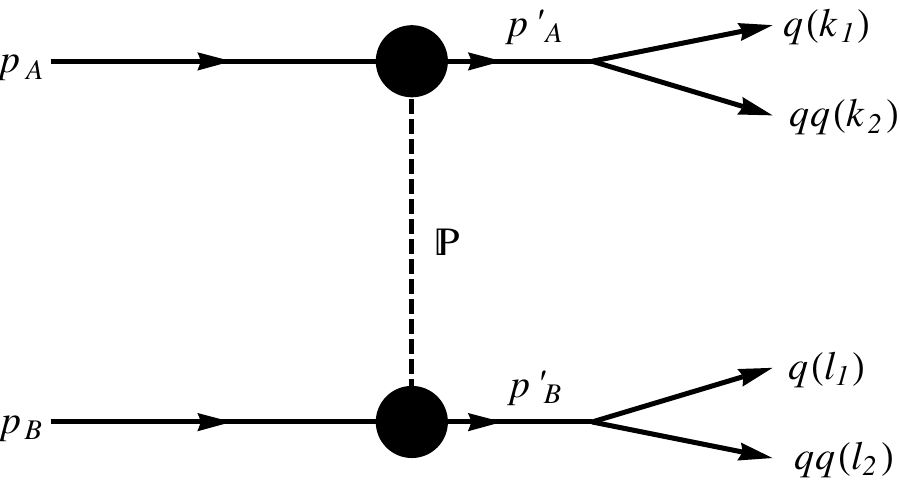}
\caption{Diffraction dissociation for single (top) and double (bottom) diffraction.}
\label{fig:diffdissoc}
\end{figure}

The implementation of diffractive dissociation in \Herwig{} is illustrated in the matrix element shown
in Fig. \ref{fig:diffdissoc} where the upper figure shows single
diffraction and the one at the bottom double diffraction. We are dealing
here with a two-to-two body problem with the incoming proton momenta
being $p_A$ and $p_B$ and the outgoing ones $p'_A$ and $p'_B$. In order
to construct the kinematics, we first sample $t$, $M_A$ and $M_B$ (for
single diffraction one of them is the proton mass $m_p$) and make sure
that one of the masses is larger. We can then compute the scattering
angle in the usual way:
\begin{align}
\cos\theta = \frac{s\left(s+2t-2m_p^2-M_A^2-M_B^2\right)}{\lambda(s,M_A^2,M_B^2)\lambda(s,m_p^2,m_p^2)},
\label{eq:scang}
\end{align}
where 
\begin{align}
\lambda(x,y,z)=x^2+y^2+z^2-2(xy+yz+xz)
\label{eq:kallen}
\end{align}
is the so-called K\"{a}ll\'{e}n function. Knowing the invariant masses
and the scattering angle it is straightforward to construct the outgoing
momenta. The dissociated proton is then decayed further into a
quark-diquark pair that moves collinear to the original hadron. This
pair in turn is converted into a cluster and taken over by the
hadronization model, where the cluster will eventually decay into two or
more hadrons.

We should point out that a fraction of the diffractive events, for very
low diffractive mass, are modeled with the $\Delta$ baryon as a final
state instead of quark-antiquark pair. Namely,
$p \ p \rightarrow \Delta \ p$ for single and
$p \ p \rightarrow \Delta \ \Delta$ for double diffraction. The $\Delta$
in turn is handled by the decay handler.  For the time being this gives
satisfactory results and is only a precautionary measure to avoid
exceptional kinematics with very light clusters.  Eventually this part
shall be taken over from the low mass end of the cluster spectrum.

\section{Soft particle production model}
\label{sec:soft}
We describe in this section the implementation of a new model for soft
interactions in \Herwig{}. With this model some of the shortcomings of
the simple model for soft interactions presented above are addressed and
the description of many minimum-bias observables is significantly
improved. 

The kinematics of soft scatterers is constructed along the lines of the
so-called multiperipheral particle production introduced in
\cite{Amati:1962nv} and we especially follow the approach taken in
\cite{Baker:1976cv}. For the case $s\gg m^2$, where $m$ is the typical
mass of a final state particle, the intermediate states depicted in
Fig. \ref{fig:ladder} via unitarity give rise to a Reggeized amplitude.
We briefly recall the main features of the intermediate state
amplitudes:
\begin{figure}[t]
\centering
\includegraphics[width=4cm]{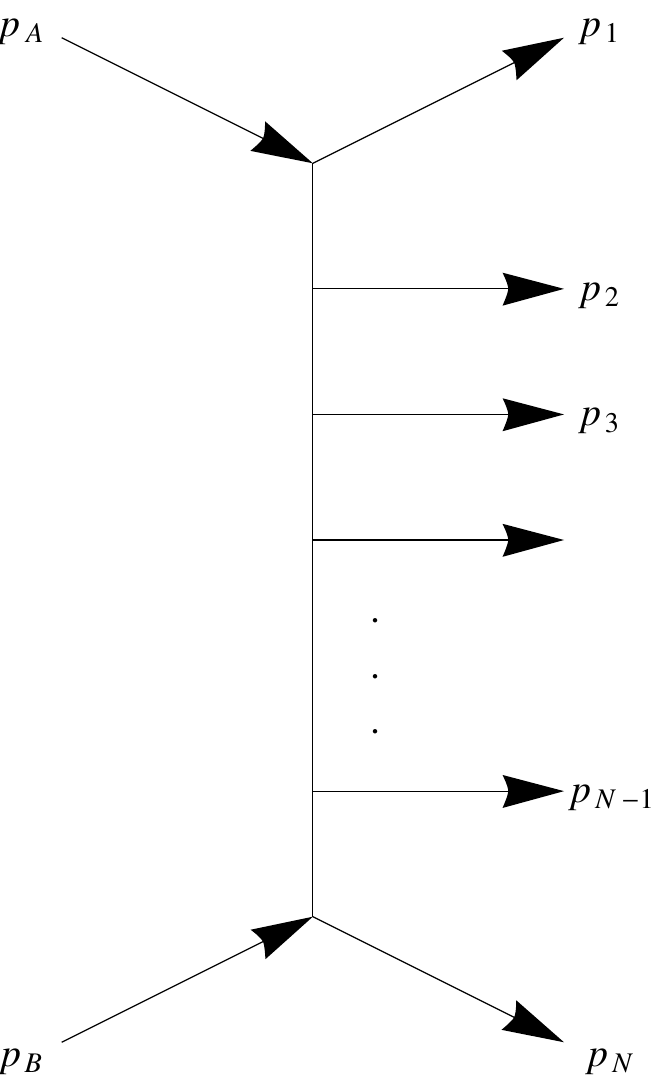}
\caption{Multiperipheral particle production.}
\label{fig:ladder}
\end{figure}
(i) The amplitude of $N$ particle production, as shown in
Fig. \ref{fig:ladder}, falls off rapidly when there is no ordering in
the longitudinal momenta for which the momentum transfer is small; (ii)
The correlation between particle momenta decreases rapidly with distance
in the ladder (i.e. momenta $p_i$ and $p_j$, where $|i-j| \gg 1$); (iii) For
sub-energies $s_{i,i+1}\equiv(p_i+p_{i+1})^2$ of the neighbouring pairs
much larger than $m^2$ the amplitude is not large. Large sub-energies
correspond to diffractive processes. Some remarks are in order. The
assumptions above lead to a fall off of the amplitude for configurations
with large rapidity separation. This gives hope that the model will fix
the so-called ``bump'' problem mentioned in the introduction. Also,
regarding point (iii), we only consider amplitudes with small
sub-energies of neighbouring particles, because we implement diffraction
using a different method as explained in the previous section.

\begin{figure}[t]
\centering
\includegraphics[width=5cm]{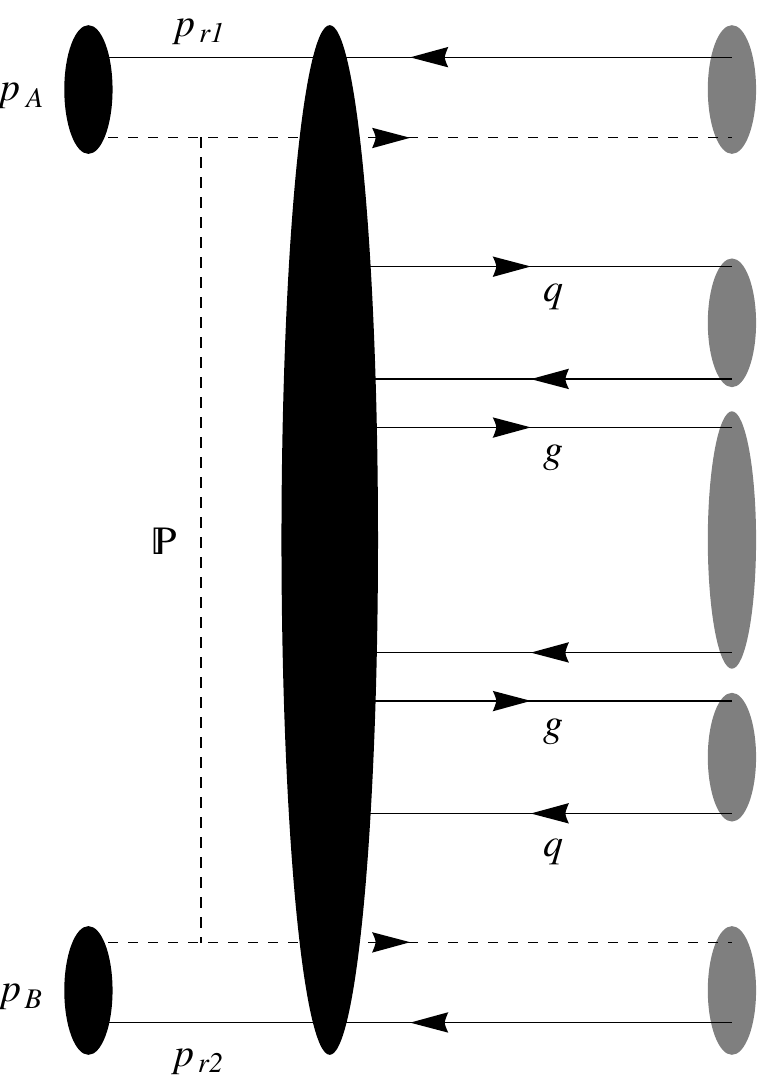}
\caption{Cluster formation in the multiperipheral final state.}
\label{fig:colconn}
\end{figure}

The model we present now uses many of the features of the old soft MPI
model, namely the eikonal model for calculating the number of soft
interactions, but implements the assumptions listed above. It should be
noted that in \Herwig{}~7, the final particles whose kinematics is
constructed using this model, will be partons, more precisely proton
remnants, sea quarks and gluons. In the following we explain the
algorithm for deriving the kinematics of final particles in more
detail. First, as in the previous versions of \Herwig{}, the soft process
starts from a quasi hard process, where a valence quark with only
longitudinal momentum is selected from the proton and the remnant takes
the rest of the momentum. This is illustrated in Fig. \ref{fig:colconn}
by the dashed lines, where a pomeron is exchanged between quasi-hard
quarks. The total energy available to perform the multiperipheral
particle production is given in the energy of the remnants. The incoming
momenta of the remnants are denoted by $p_{r1}$ and $p_{r2}$ in
Fig. \ref{fig:colconn}. According to \cite{Baker:1976cv} the number $N$
of the final particles in the ladder is drawn from a Poissionian
distribution with mean
\begin{align}
	\avg{N} = n_{\rm ladder} \ln \frac{(p_{r1}+p_{r2})^2}{m_{\mathrm{rem}}^2},
\label{eq:avgN}
\end{align}
where $m_{\mathrm{rem}}$ is the constituent mass of the remnant and
$n_{\rm ladder}$ is a constant which is very close to one and will be tuned
below to minimum-bias data. Fig. \ref{fig:colconn} illustrates a case
with $N=6$, where we have two remnants, a sea quark and an antiquark and
two gluons.

In the following we adopt the algorithm described in \cite{Baker:1976cv}
for generating the kinematics of final state partons, which give
diagrams with amplitudes satisfying the assumptions above. The momenta
are separated into their longitudinal and transverse parts, namely
\begin{align}
p_i=(p_{0i}, \boldsymbol{p}_{i\perp}, p_{iz}).
\label{eq:momsep}
\end{align}
 It was shown in \cite{Baker:1976cv} that for the case
\begin{align}
p_{iz}^2 \gg m_i^2+p_{i\perp}^2
\label{eq:longmom}
\end{align}
where $p_i^2=m_i^2$, and assuming the same holds for momentum transfer
between neighboring elements in Fig. \ref{fig:ladder}, then the
longitudinal momenta can be generated by the following rule:
\begin{align}
&p_{1+}=x_1 p_{r1}, \ \ p_{2+}=(1-x_1)x_2 p_{r1}, \ldots, \nonumber \\&p_{i+}=(1-x_1)(1-x_2)\cdots(1-x_{i-1})x_ip_{r1}.
\label{eq:lcmom}
\end{align}
where $p_{i\pm}\equiv p_{i0}\pm p_{iz}$ and $x_i$ take values between
$0$ and $1$. We want to ensure partons are separated equally in
rapidity. We assume all $x_i\approx x$ to have roughly the same
value. Consider the total rapidity between remnants $\Delta Y$. The
spacing in rapidity between partons, after the number $N$ is sampled, is
\begin{align}
\Delta y  = \frac{1}{N-1}\Delta Y\simeq \ln\frac{p_i}{p_{i-1}}.
\label{eq:dy}
\end{align}
Using \eqref{eq:dy} and \eqref{eq:lcmom}, we can compute:
\begin{align}
x=1-e^{-\Delta y}=1-e^{-\frac{1}{N-1}\Delta Y}.
\label{eq:xavg}
\end{align} 
Longitudinal momenta are thus generated from \eqref{eq:xavg} and
\eqref{eq:lcmom}. Transverse momenta are sampled from
\eqref{eq:gaussian}.  In order to facilitate the proper colour flow, we
have to introduce a pair of quark-antiquark as shown in
Fig. \ref{fig:colconn}.

Let us explain in a bit more detail the picture in
Fig. \ref{fig:colconn}. The initial quark extracted from the proton is
colour connected with the remnant and form a cluster (clusters are
denoted by gray blobs in the figure). The same holds on the other end
for the other proton. The sea quark, denoted by $q$ is colour connected
to the first gluon, denoted by $g$. The subsequent gluons are connected
with their neighbours. The same holds for the other proton where instead
of a quark, we have an antiquark (also denoted by $q$). Since the first
quark is extracted from the proton using a parton distribution function
(PDF), we have to make sure that its rapidity is close to the rapidity
of second particle in the ladder, which in our case is the sea
quark. This can be done by choosing the proper value of
$x_{\mathrm{min}}$ of this PDF.

The algorithm presented in this section guarantees exponential fall off
of the amplitude for large values of rapidity separation $\Delta \eta$.
Also, it gives a roughly flat distribution in rapidity of the clusters
and the subsequently produced particles.

\begin{figure}[t]
\centering
\includegraphics[width=7cm]{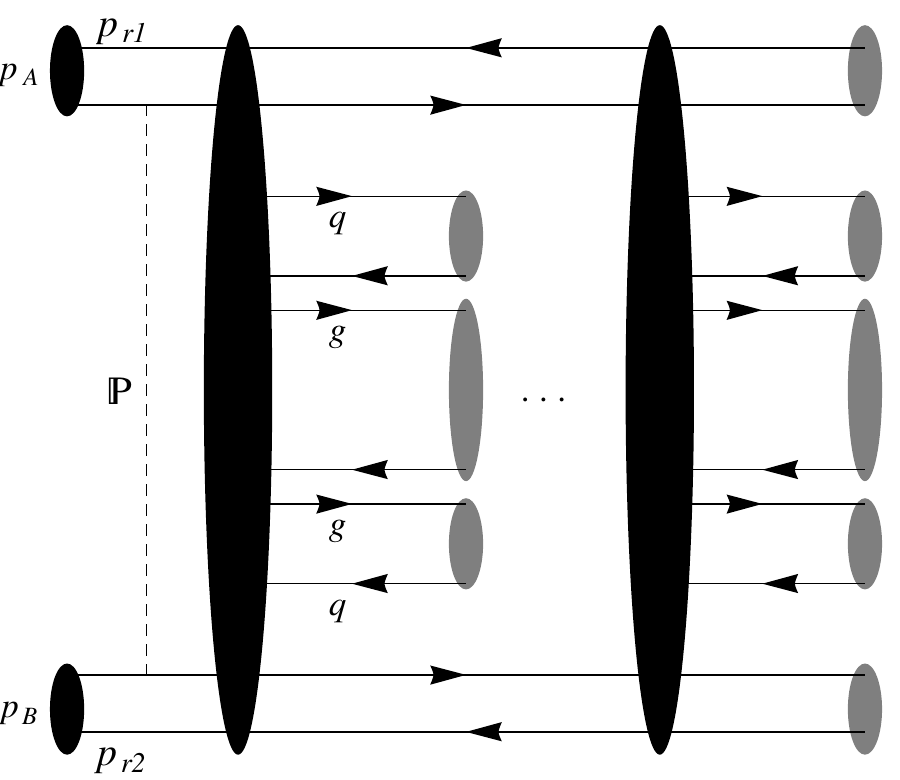}
\caption{Cluster formation in the multiperipheral final state with multiple interactions.}
\label{fig:colconn_mult}
\end{figure}

Finally, it should be noted that we take into account also soft multiple
parton interactions.  This would correspond to intermediate amplitudes
with many multiperipheral final states for a given event. The
probability for having $k$ soft interactions is computed from the
existing model in Herwig (see \cite{Bahr:2008pv}). The implementation of
such a final state is shown in Fig. \ref{fig:colconn_mult}.

\section{Tuning}
\label{sec:tuning}
In order to model minimum-bias one must also include single and double
diffractive events which makes minimum-bias modeling more complicated
than the modeling of the UE. With the new model for
diffraction and soft particle interaction the new model can be tuned to
fit minimum-bias data.  In this section we describe the tuning of the
new model to data from hadron colliders. The main part of the tuning is
achieved by using the \PROFESSOR{} framework
\cite{Buckley:2009bj}. Since we changed the soft part of the MPI model
we need to re-tune all parameters that affect this model. The main
parameters of the MPI model are the parameters of the
$\ptmin$ parametrization, $p_{\perp,0}^{\rm min}$ and
$b$, presented in Ref. \cite{Gieseke:2012ft} and the inverse proton
radius squared $\mu^2$. Also considered in the tuning is the colour
reconnection probability $p_{\mathrm{reco}}$ and the only new parameter
of the model, the ladder multiplicity $n_{\rm ladder}$ introduced in expression \eqref{eq:avgN} above.  At the same
time we get rid of the parameter $P_{\rm disrupt}$ from the old soft
interaction model, that described the probability of choosing a
disrupted colour connection.  Hence, in total we keep the number of
tunable parameters fixed with introducing the new model for soft
interactions.

The parameters governing hadronization were tuned to LEP data
\cite{Bahr:2008pv} and are left untouched.  We tune the model to minimum
bias data from the ATLAS collaboration at
$ \sqrt{s}=900 \, \mathrm{GeV}$ and $\sqrt{s}=7 \, \mathrm{TeV}$
\cite{Aad:2010ac}.  For the tuning procedure we use the following eight
observables with equal weights:
\begin{itemize}
	\item[•] The pseudorapidity distributions for\\ $N_{\mathrm{ch}} \geq 1$, $N_{\mathrm{ch}} \geq 2$ , $N_{\mathrm{ch}} \geq 6$, $N_{\mathrm{ch}} \geq 20$,
	\item[•] The transverse momentum of charged particles for\\  $N_{\mathrm{ch}} \geq 1$, $N_{\mathrm{ch}} \geq 2$ , $N_{\mathrm{ch}} \geq 6$,
	\item[•] The charged transverse momentum vs. number of\\ charged particles for $N_{\mathrm{ch}}\geq 1$.
\end{itemize}
For the tuning of 5 parameters with a 4-dimensional interpolation we
generate 500 runs consisting of 500000 events each with randomly
selected parameter values within a specified range. A subset of these
500 runs is then used 350 times in order to interpolate the generator
response. This also serves as a cross check if the interpolation does
indeed find the minimum value. For each of these run combinations the
$\chi^2/N_{\mathrm{dof}}$ is calculated and real Monte Carlo runs were
performed in order to verify if the interpolation did predict the right
value of $\chi^2/N_{\mathrm{dof}}$. The set of parameters that resulted
in the smallest value of $\chi^2/N_{\mathrm{dof}}$ was then used for
further analyses. 

The tuning to minimum bias data resulted in two slight\-ly different sets
of parameters for $\sqrt{s}=900 \, \mathrm{GeV}$ and
$\sqrt{s}=7000 \, \mathrm{GeV}$.  The 7000\,GeV tune will serve as the
default minimum bias tune for now. We note that the parameters of the
$\ptmin$ parametrization have approximately the same value
in both tunes, which indicates that the parametrization is stable with
respect to energy extrapolation, which gets confirmed with our runs at
13\,TeV.

The new model with the tuned parameters clearly improves the description
of all observables which were considered in the tuning itself.  This
will be shown in the next section.

\section{Results}
\label{sec:results}
\subsection{Rapidity Gap Analysis}

In Refs.~\cite{Aad:2012pw} and \cite{Khachatryan:2015gka} the
differential cross section with respect to the forward pseudorapidity
gap $\Delta \eta^F$ is measured. $\Delta \eta^F$ is defined as the
larger of the two pseudorapidity regions extending to the boundary of the
detector in which no particles are produced. The acceptance in
pseudorapidity $\eta$ ranges from $-4.9$ to $+4.9$ at ATLAS and from
$-4.7$ to $+4.7$ at CMS, which is restricted by the geometry of the
detectors. All particles with $p_{\perp}>p_{\perp}^{\mathrm{cut}}$ are
analyzed where $p_{\perp}^{\mathrm{cut}}$ is varied from
$200 \, \mathrm{MeV}$ to $800 \, \mathrm{MeV}$. The total cross section
is usually decomposed into the non-diffractive (ND),
single/double-diffractive dissociation (SD/DD) and central-diffractive
(CD) parts. The latter is suppressed with respect to other
contributions. Events with small pseudorapidity gaps are mainly
dominated by ND contributions and for a small $p_{\perp}^{\mathrm{cut}}$
the large rapidity gap region is dominated by SD and DD events.  The ND
part is characterized by the experimental observation that the average
rapidity difference between neighbouring particles is around 0.15 with
larger rapidity gaps due to fluctuations in the hadronization
process. This leads to a cross section that decreases exponentially with
larger rapidity gaps $\sigma_{ND} \sim \exp(-a\Delta \eta^F)$ where $a$
is some constant. Events with large pseudorapidity gaps,
dominated by diffractive events which result from pomeron exchange as
briefly reviewed above, at large energies give rise to a constant cross
section $\sigma_{D} \approx \mathrm{const}.$ in $\Delta \eta^F$.

\begin{figure*}[t]
  \centering
  \includegraphics[width=0.49\textwidth]{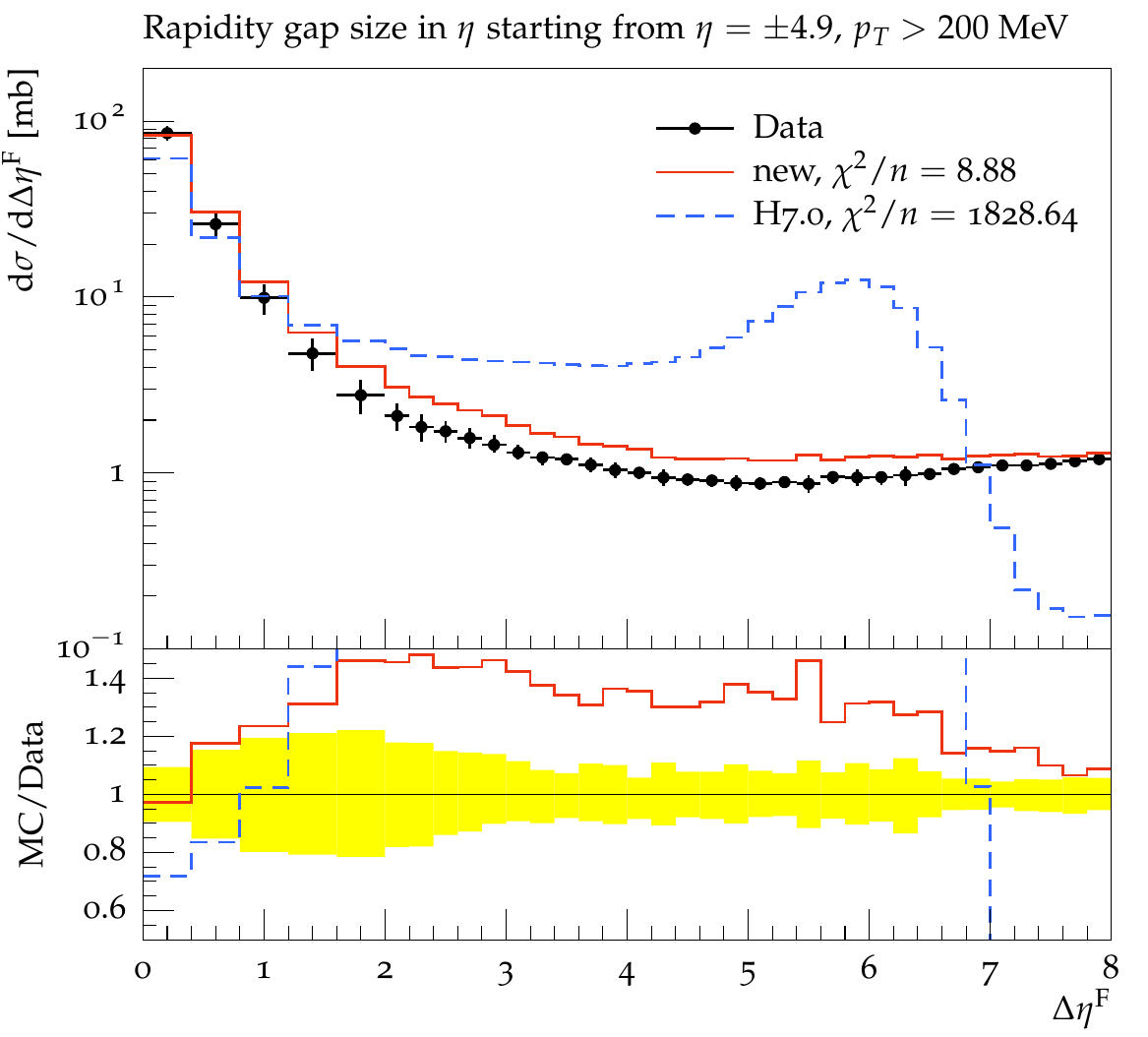}\hfill
  \includegraphics[width=0.49\textwidth]{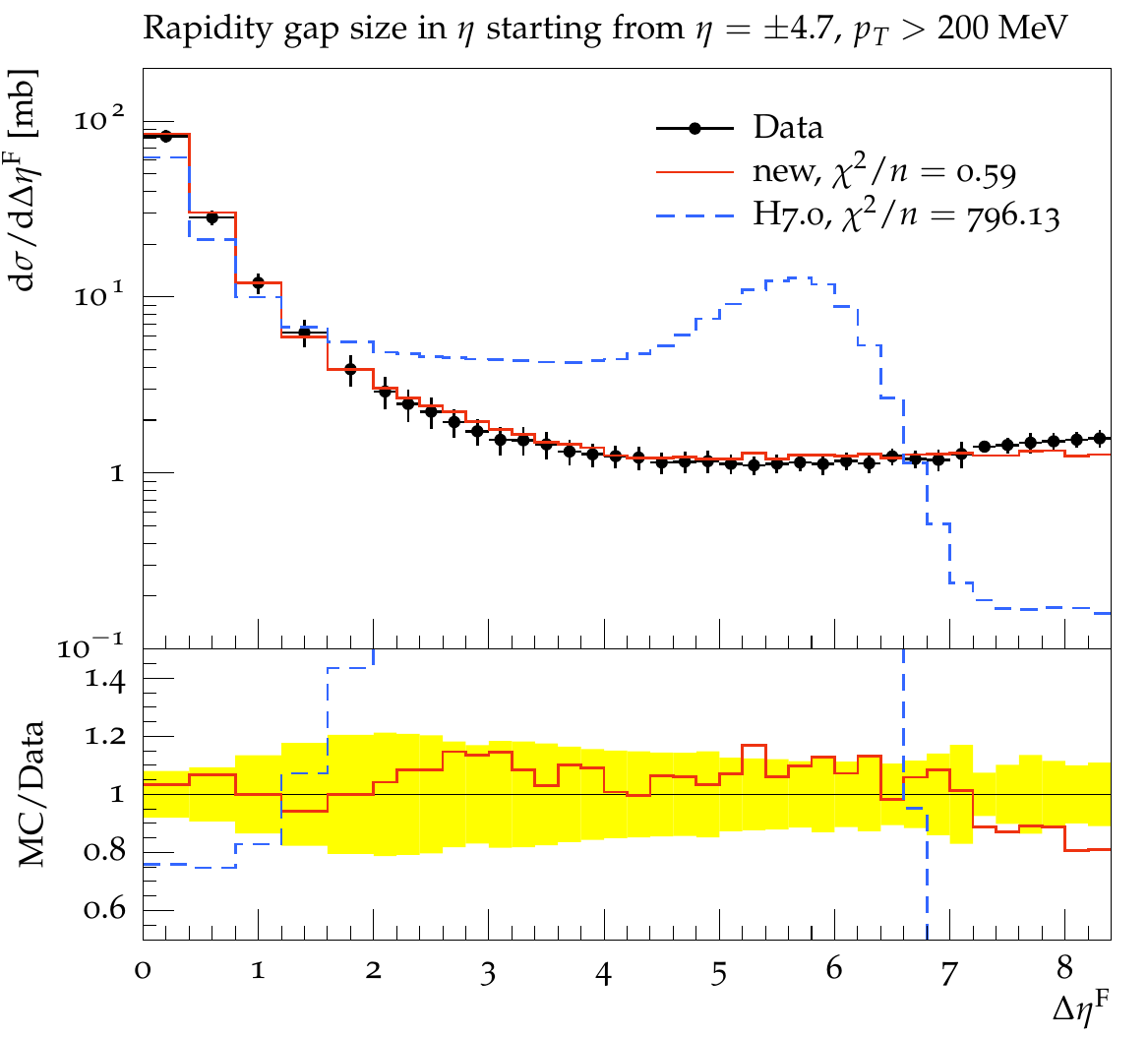}
  \caption{Comparison of the new model for soft interactions and
    diffraction with the old model from \Herwig{}~7 to ATLAS rapidity gap
    measurement at $\sqrt{s} = 7 \, \mathrm{TeV}$ with
    $p_{t}>200\, \mathrm{MeV}$.  In the left panel we copare to ATLAS
    data in the range $|\eta| \leq 4.9$ \cite{Aad:2012pw}.  In the right
    panel we compare to data from CMS in the range $|\eta|\leq 4.7$
    \cite{Khachatryan:2015gka}. }
  \label{fig:gaps}
\end{figure*}

By combining the model for the simulation of diffractive events, as
reviewed in section \ref{sec:diffraction}, with the new model for soft
particle production proposed in \ref{sec:soft}, we can describe quite
well the measurement of the rapidity gap cross section from ATLAS
\cite{Aad:2012pw} and CMS \cite{Khachatryan:2015gka}. Results for
$p_{\perp}>200$\,MeV are shown in Fig.~\ref{fig:gaps}. It should be noted that
while the data from CMS is described very well, the simulation
overestimates the data provided by ATLAS despite quite similar cuts.

\subsection{Minimum-bias data}
\begin{figure*}[p]
  \begin{minipage}{0.5\textwidth} 
    \includegraphics[width=\textwidth]{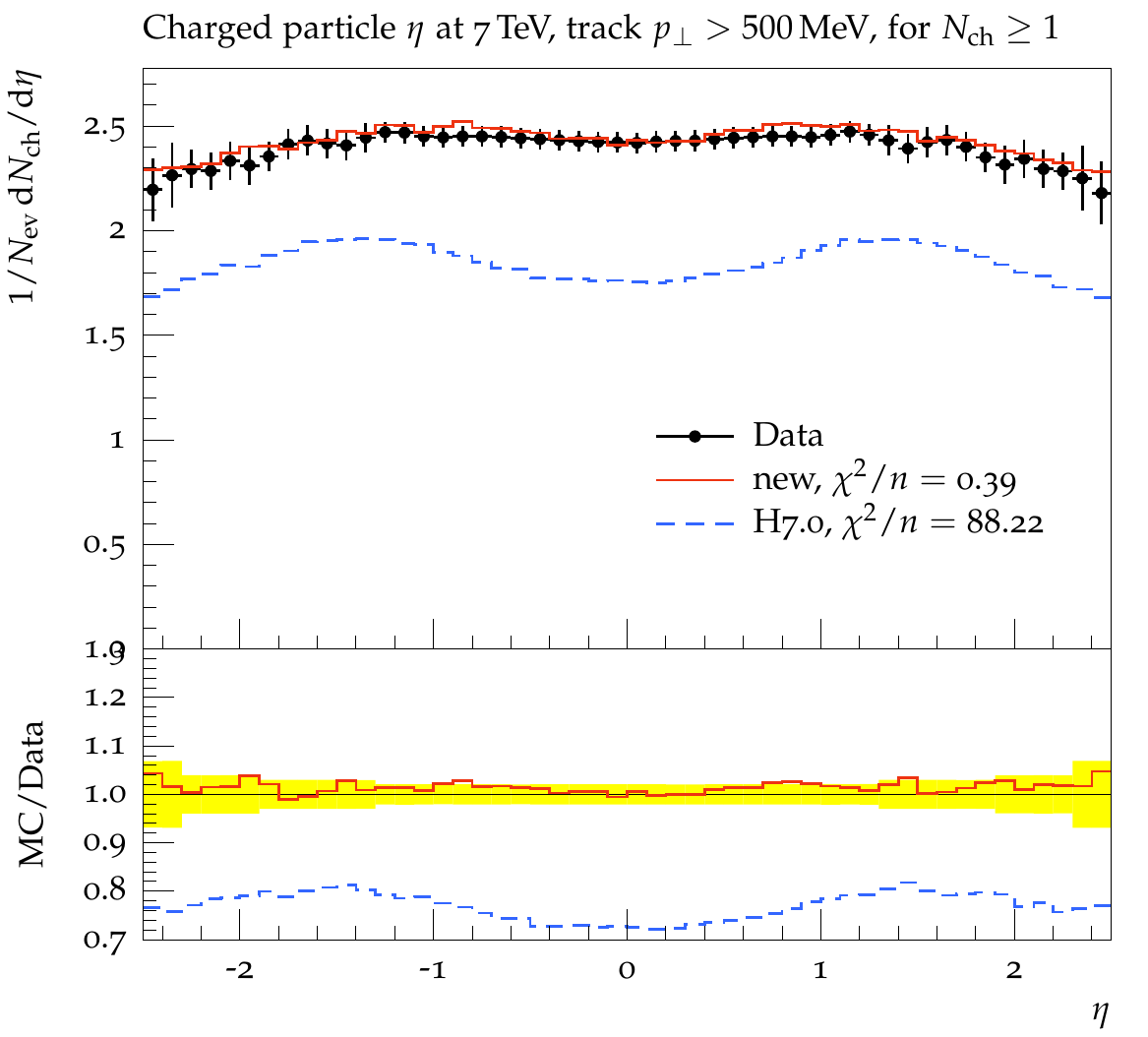}
  \end{minipage}
  \hfill
  \begin{minipage}{0.5\textwidth}
    \includegraphics[width=\textwidth]{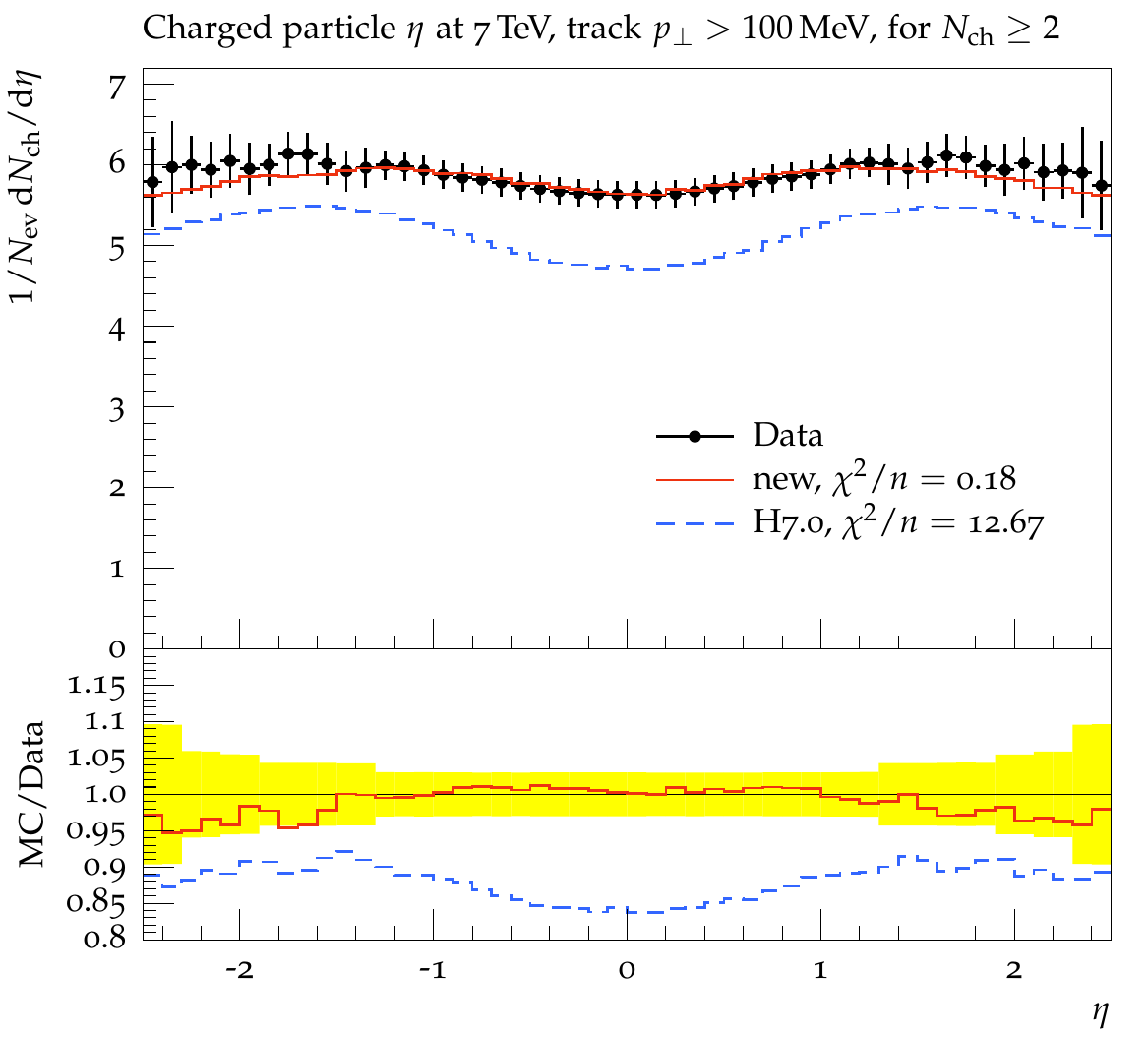}
  \end{minipage}
  \begin{minipage}{0.5\textwidth} 
    \includegraphics[width=\textwidth]{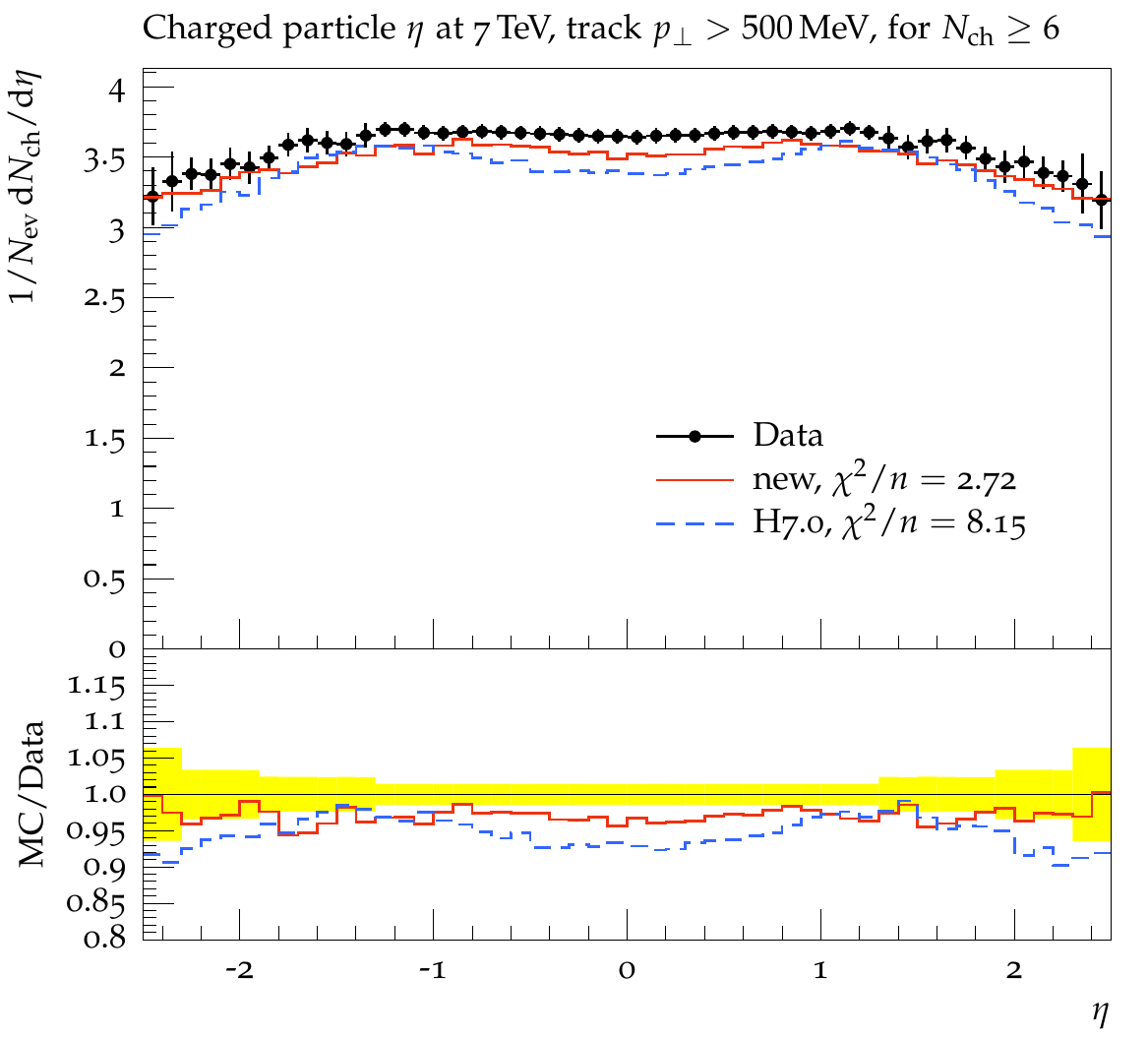}
  \end{minipage}
  \hfill
  \begin{minipage}{0.5\textwidth}
    \includegraphics[width=\textwidth]{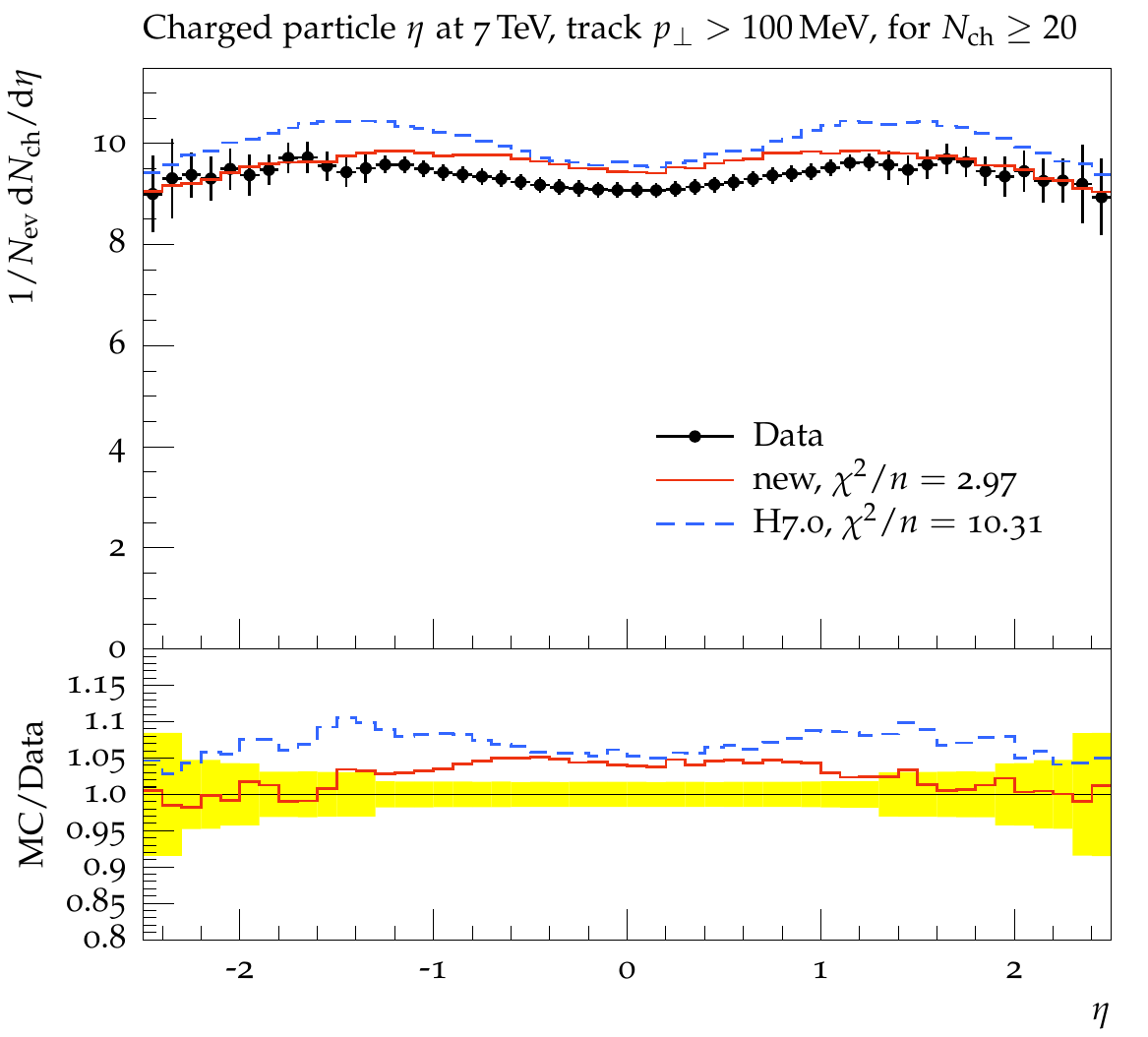}
  \end{minipage}	
  \caption{Comparison of the default tune from Herwig~7.0 with the new
    model to minimum-bias data from ATLAS \cite{Aad:2010ac} at
    $\sqrt{s} = 7 \, \mathrm{TeV}$.}
 \label{fig:MB7}
\end{figure*}

\begin{figure*}[p]
\begin{minipage}{0.5\textwidth}
  \includegraphics[width=\textwidth]{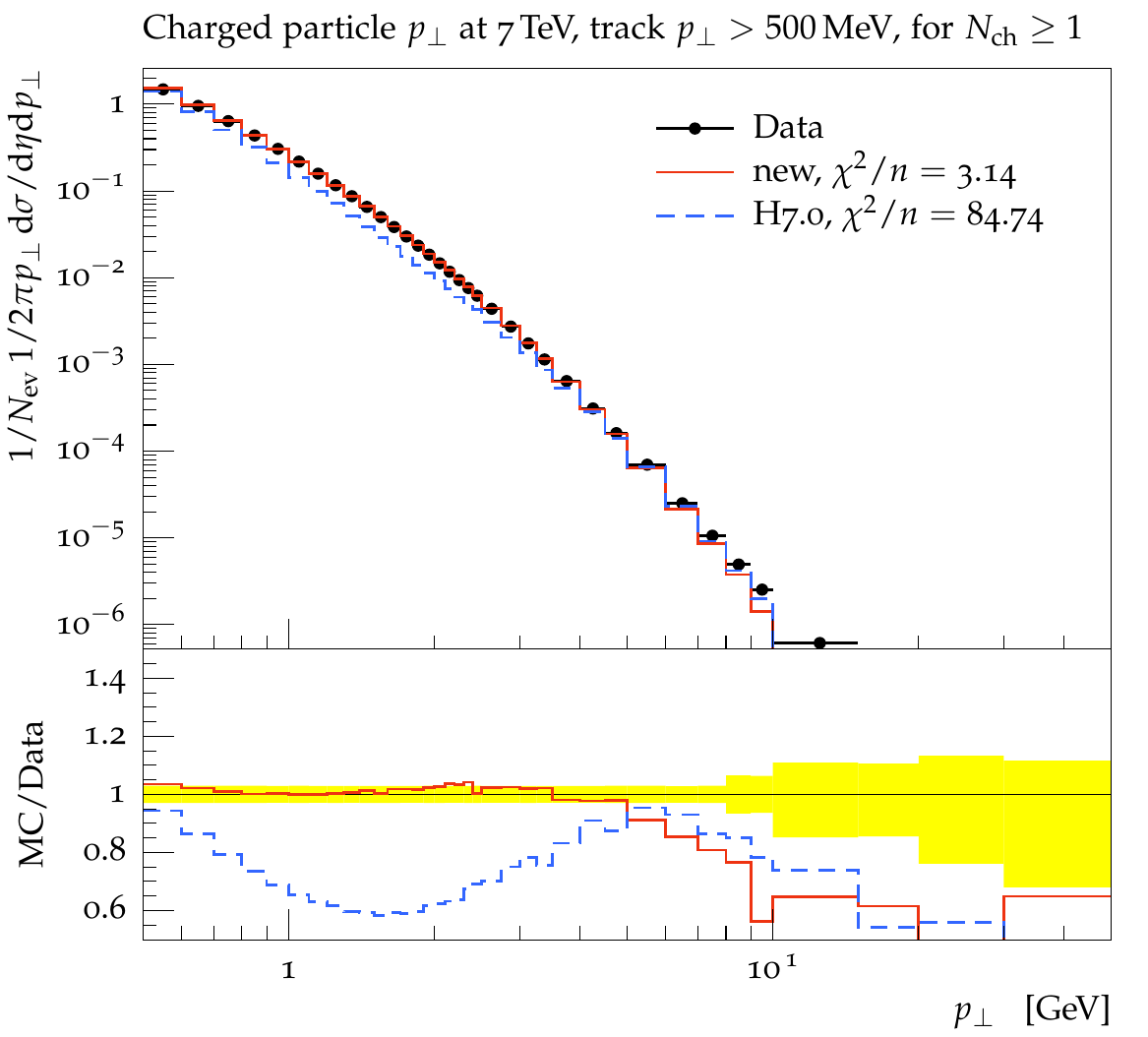}
\end{minipage}
\begin{minipage}{0.5\textwidth}
  \includegraphics[width=\textwidth]{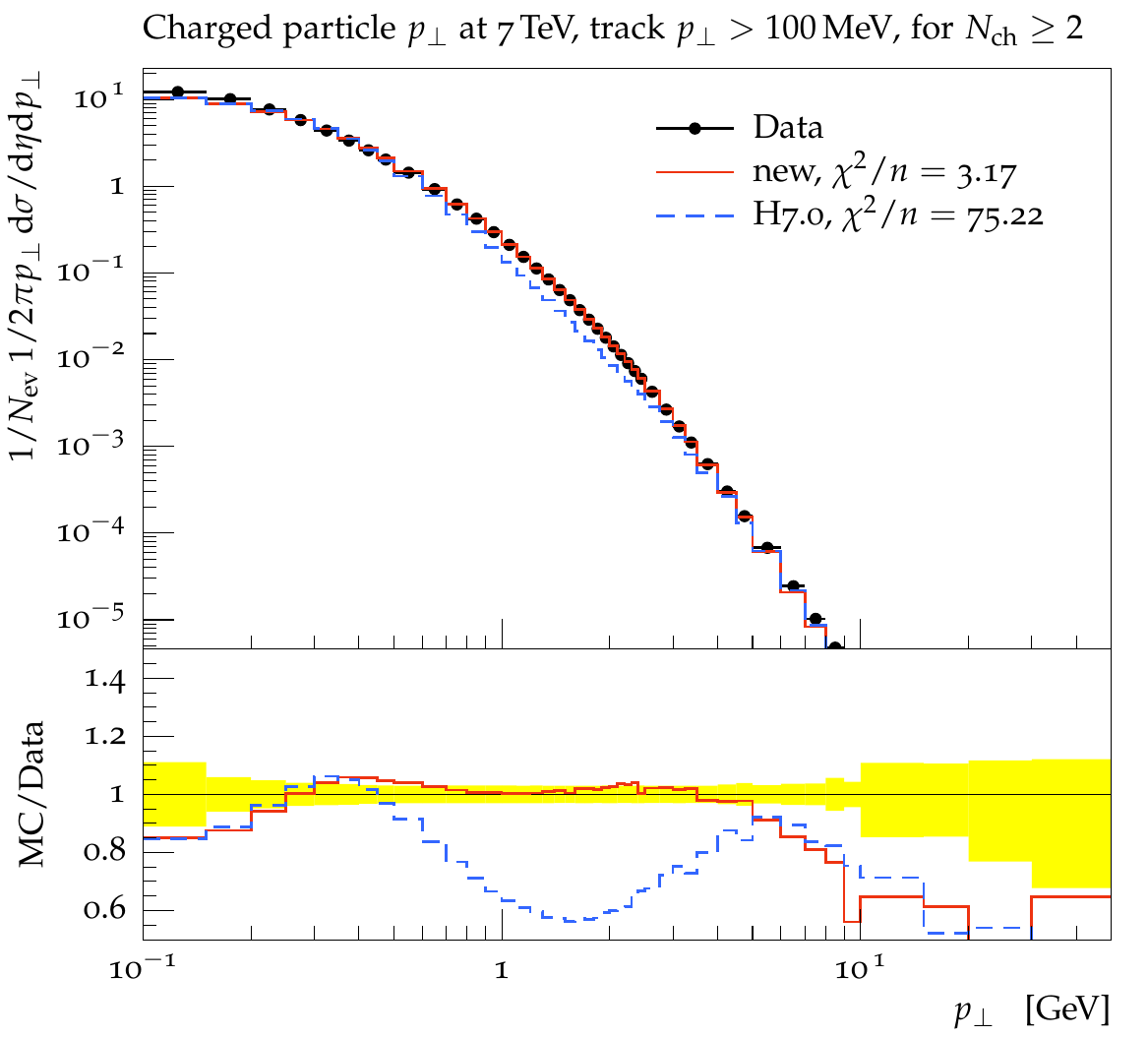}
\end{minipage}

\caption{Comparison of the default tune from Herwig~7.0 with the new
  model to minimum-bias data from ATLAS \cite{Aad:2010ac} at
  $\sqrt{s} = 7 \, \mathrm{TeV}$.}
 \label{fig:MB7-1}
\end{figure*}

We further test the new model versus many different mini\-mum-bias
measurements. Most observables are signifi\-cant\-ly improved, although we
only tuned to a small subset of available observables. The results for
the Monte Carlo runs with the tuned parameters for $7\, \mathrm{TeV}$
are shown in Fig.~\ref{fig:MB7}. Here we show all $\eta$ distributions,
and we notice that the overall description is quite good. The
distribution for the charged particle $p_{\perp}$ versus the number of
charged particles is shown in Fig. \ref{fig:MB7-1}, for different
cuts. We notice that for small $p_\perp$ cut the model fails to describe
the data.  This observable is very sensitive to models of colour
reconnection that add correlations to final state particles from
previously uncorrelated events of MPI models
\cite{Sjostrand:1987su}. Overall, it is especially noteworthy to mention
that the new model fits the charged particle $p_{\perp}$ distribution
almost perfectly in the range where we expect it to contribute
significantly. Also the onset of the charged particle $p_{\perp}$ versus
the number of charged particles improves which is due to
diffraction. The tail of this distribution seems to underestimate the
$p_{\perp}$ value but the tune results in an overall better description
of the observables.

\begin{figure*}
\begin{minipage}{0.5\textwidth}
	\includegraphics[width=\textwidth]{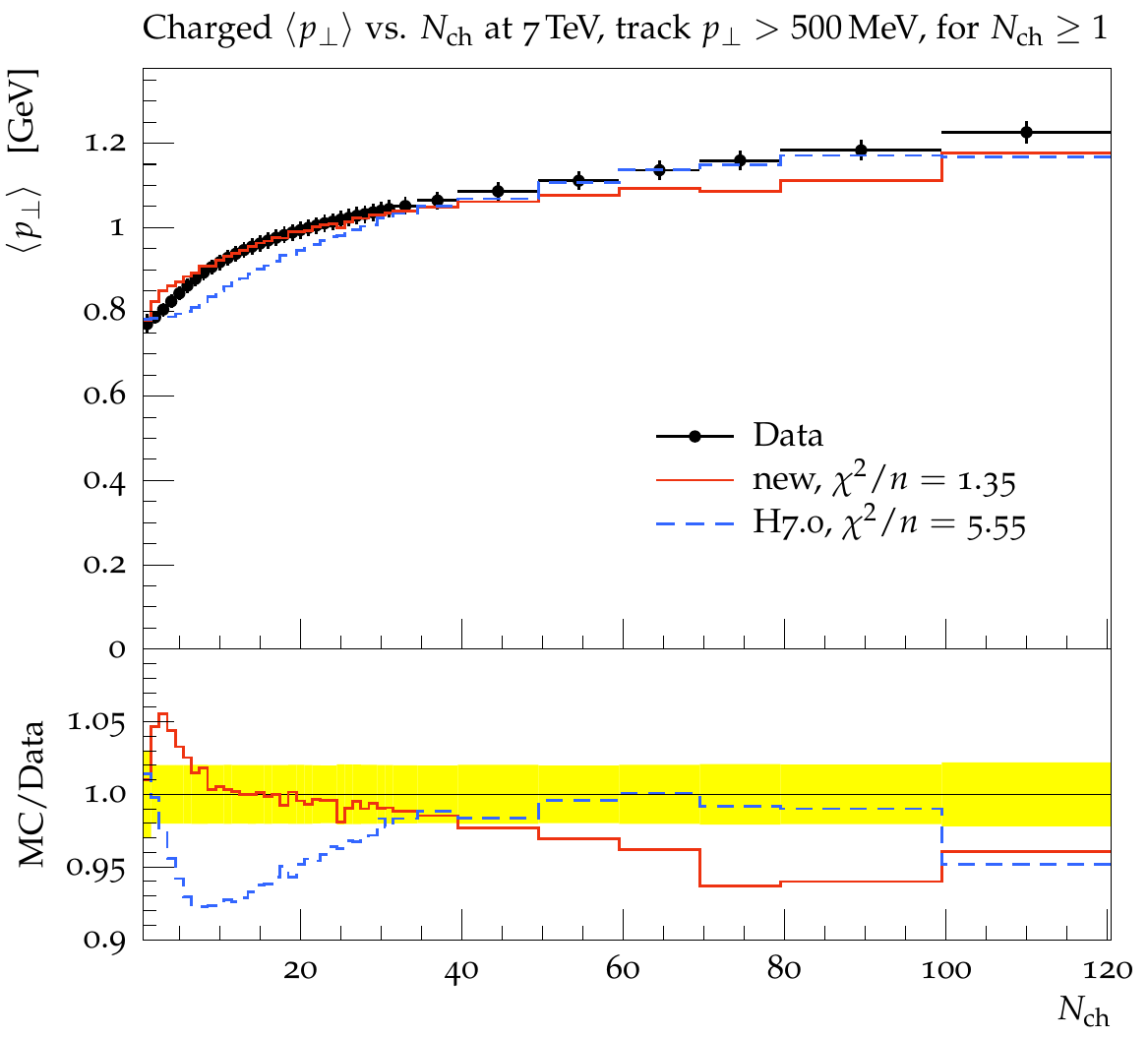}
	\end{minipage}
	\begin{minipage}{0.5\textwidth}
	\includegraphics[width=\textwidth]{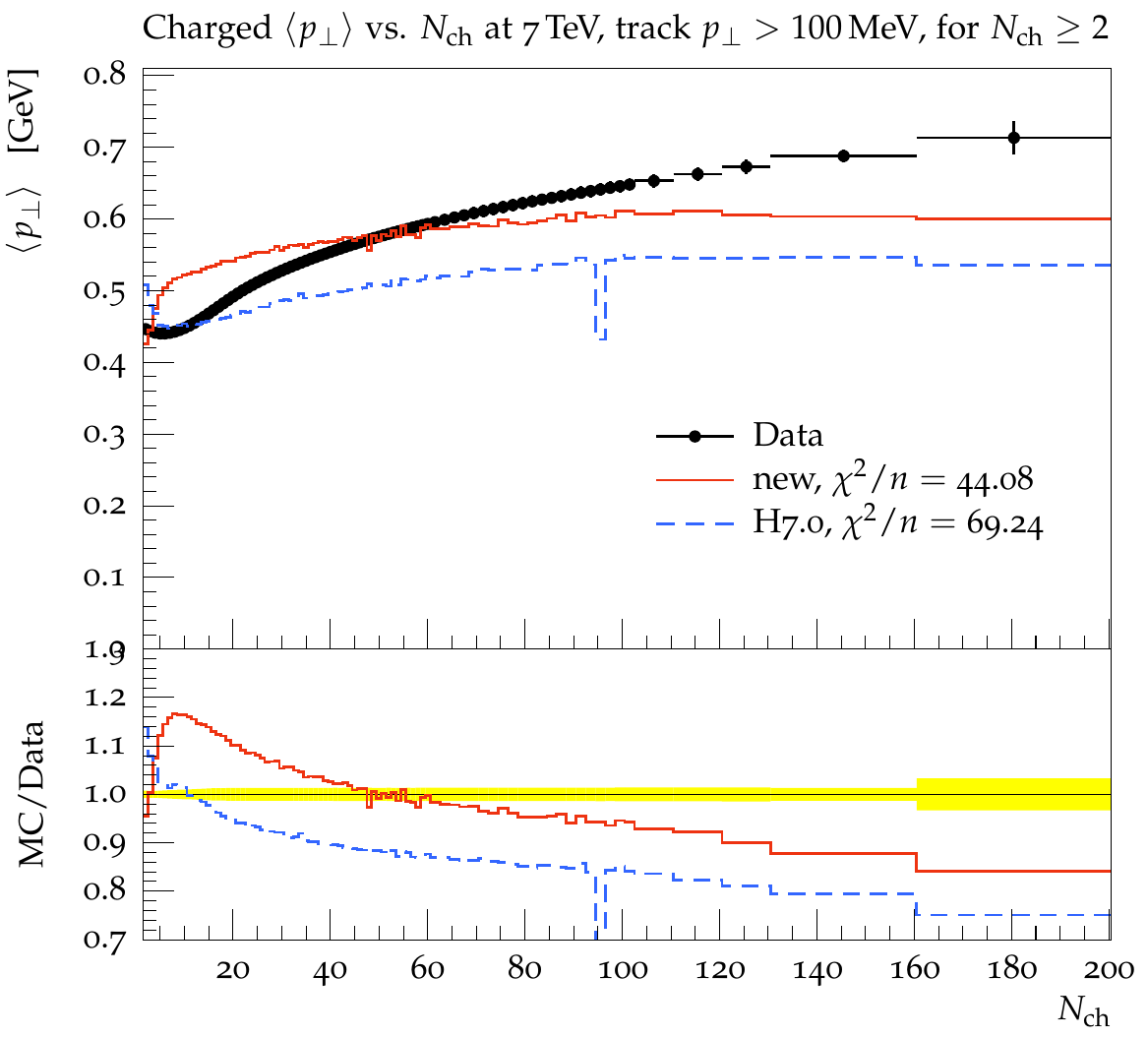}
	\end{minipage}
	\caption{Comparison of the default tune from \Herwig{}~7 with the new model to minimum-bias data from ATLAS \cite{Aad:2010ac} at $\sqrt{s} = 7 \, \mathrm{TeV}$.}
	\label{fig:MB7-pp}
\end{figure*}

In Fig. \ref{fig:MB7-pp} the average charged particle $p_{\perp}$ versus the
number of charged particles $N_{\rm ch}$ is shown. While there is an improvement
for low $N_{\rm ch}$ due to perhaps the diffractive model, the overall result is
unsatisfactory for very soft particles ($p_\perp > 100\,$MeV).  Here,
the colour reconnection of soft particles appears to have too small an
effect in order to result in a rise of the this observable for larger
$N_{\rm ch}$.  We note that the transverse momentum spectra of charged
particles are much improved in our model with respect to our old model.
This is interesting, as sometimes a failure to describe these spectra in
older models has been attributed to a lack of collective effects.

\subsection{Analysis of non-single-diffractive events}

The analysis presented in Ref.~\cite{Khachatryan:2010nk} is based on an
event selection which is corrected according to the SD, DD and ND events
predicted by \Pythia~6 \cite{Sjostrand:2006za}. Therefore this analysis
is automatically biased by these predictions. It is nonetheless useful
to see how our new model performs with respect to these observables.

\begin{figure*}[p]
  \begin{minipage}{0.5\textwidth} 
    \includegraphics[width=\textwidth]{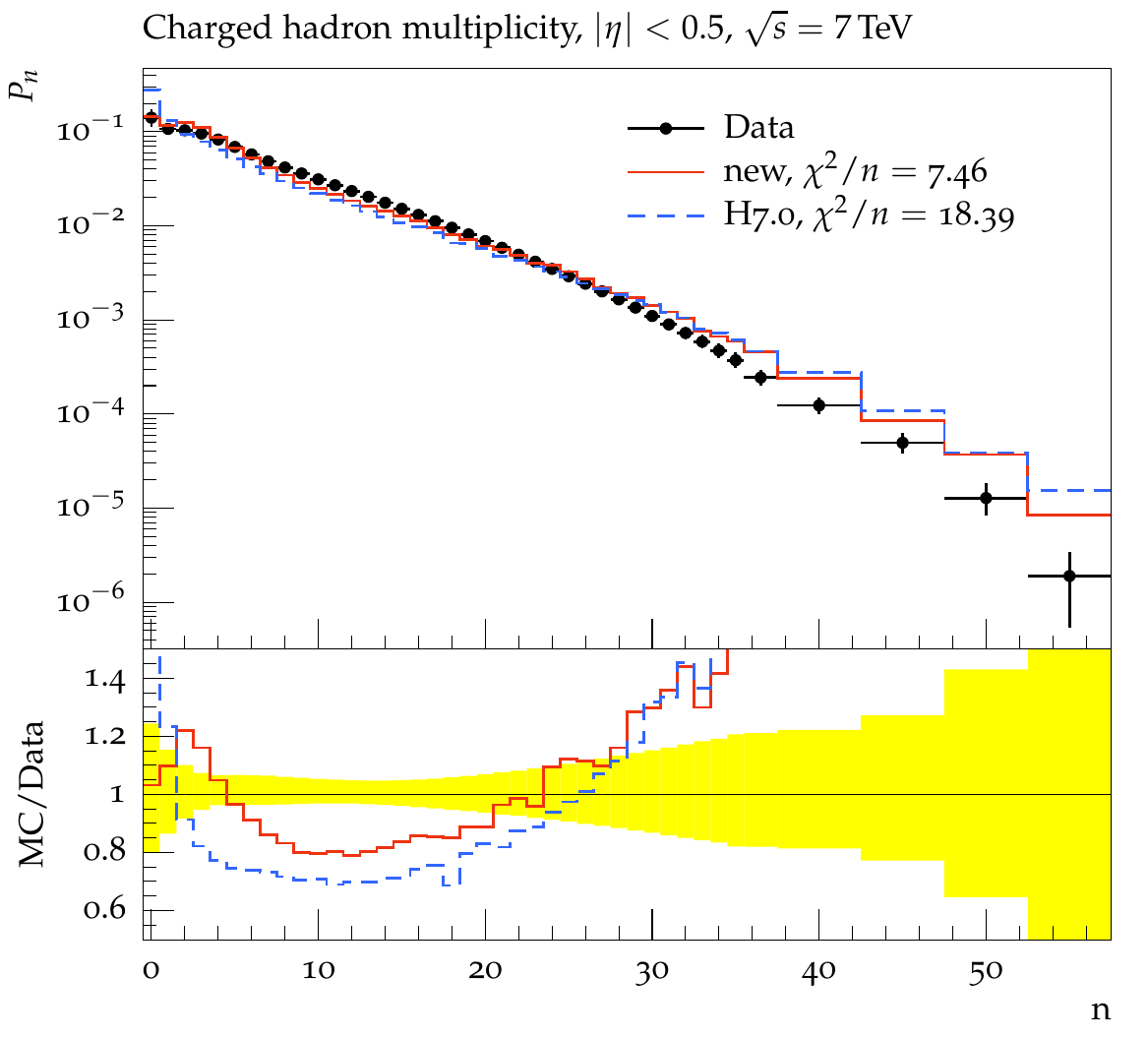}
  \end{minipage}
  \hfill
  \begin{minipage}{0.5\textwidth}
    \includegraphics[width=\textwidth]{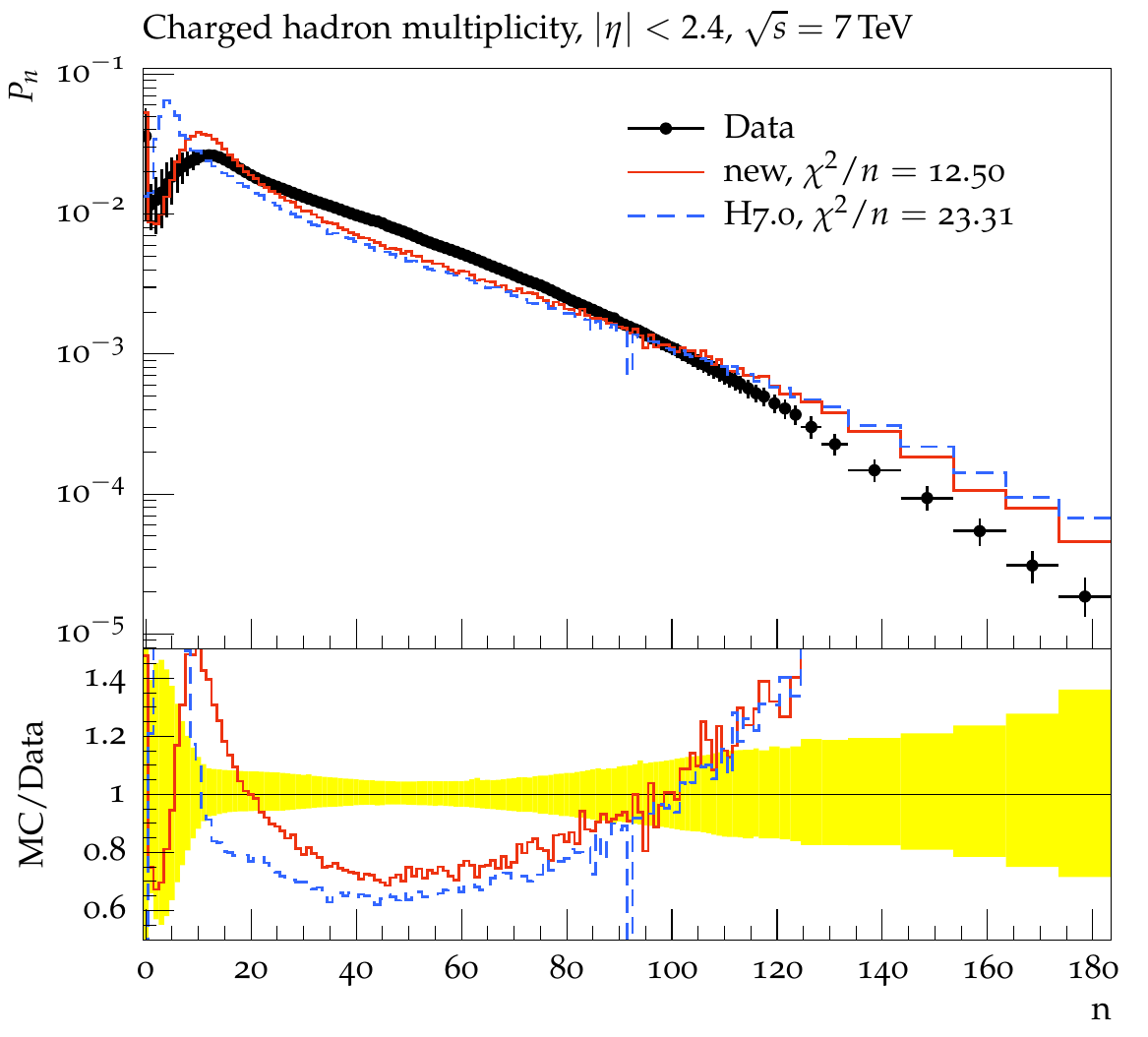}
  \end{minipage}
  \caption{Multiplicity distributions for the very central region
    $|\eta|<0.5$ (a) and the most inclusive measurement by CMS
    \cite{Khachatryan:2010nk} (b). In order to compare to the data only
    non-single-diffractive events were simulated with the new model
    while H7.0 uses the old model for MPI and lacks a model for
    diffraction completely.}
  \label{fig:NSD1}
\end{figure*}

Although we note significant improvement in the region of low
multiplicity the new model fails to describe the data correctly (see
Fig.~\ref{fig:NSD1}). It is interesting to note that in
Ref.~\cite{Khachatryan:2010nk} it was found that the event generators
systematically underestimated the increase of the multiplicity
distribution while our model (and also the old default model)
overestimate it.  The multiplicity distribution is mainly influenced by
the mass distribution of the clusters. The higher the cluster mass, the
more particles get produced from the cluster. We expect a change in the
colour reconnection model to have significant impact on these
distributions which will be studied in more detail in the near future.

\begin{figure*}[p]
  \begin{minipage}{0.5\textwidth} 
    \includegraphics[width=\textwidth]{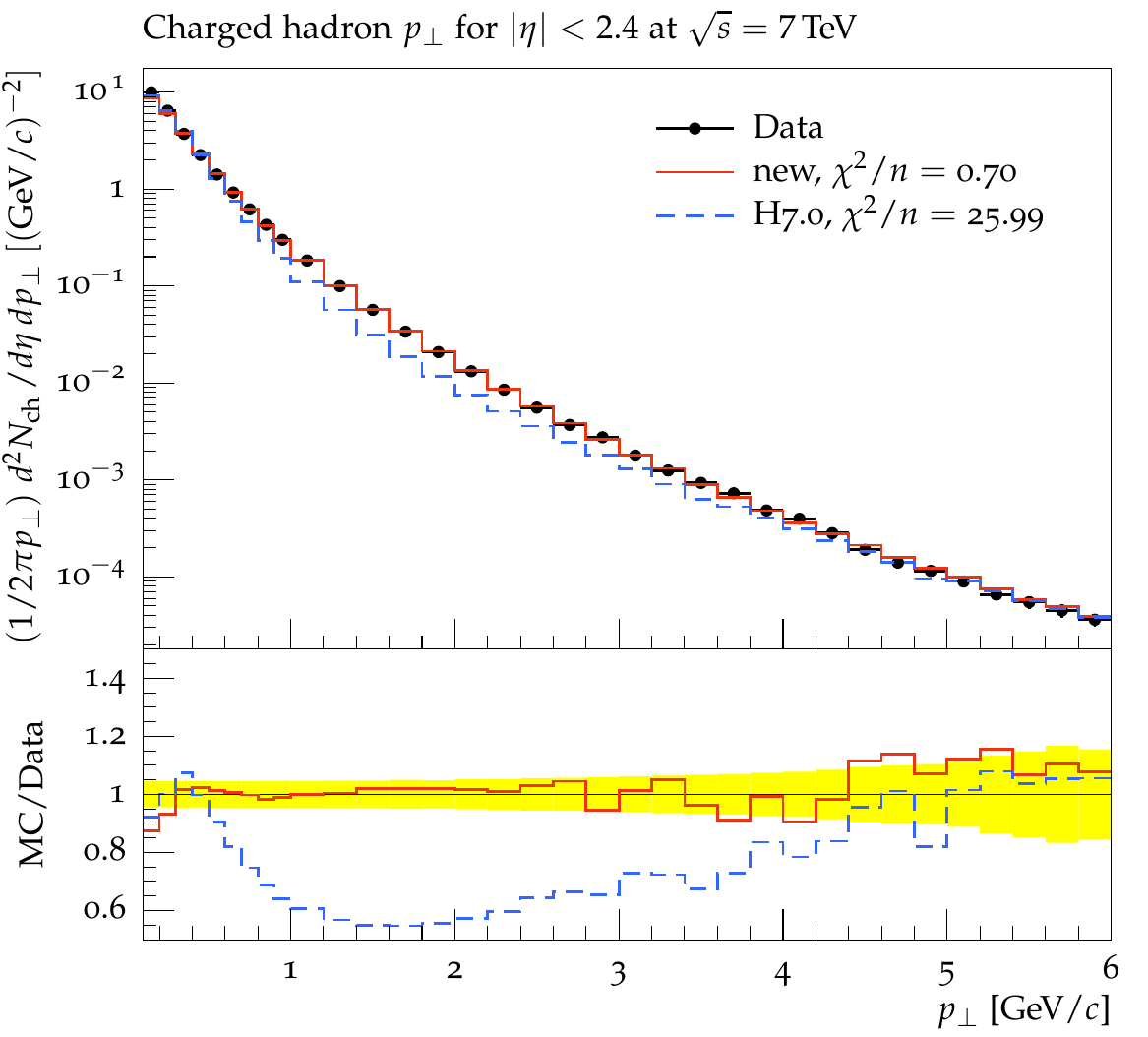}
  \end{minipage}
  \hfill
  \begin{minipage}{0.5\textwidth}
    \includegraphics[width=\textwidth]{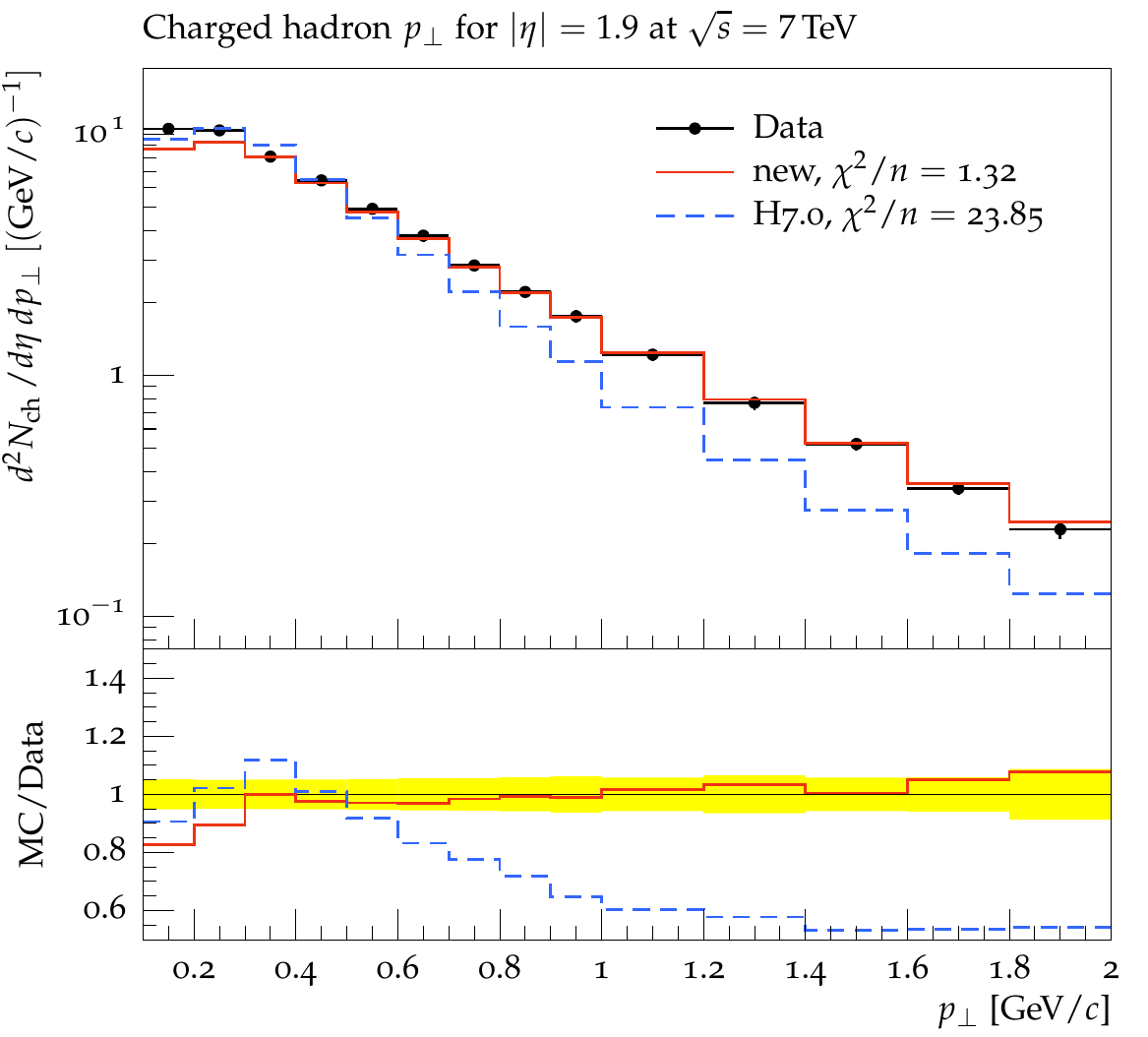}
  \end{minipage}
  \caption{$p_{\perp}$ distributions for $|\eta|<2.4$ and $|\eta|=1.9$
    measured by CMS \cite{Khachatryan:2010us}. In order to compare to
    the data only non-single-diffractive events were simulated with the
    new model while H7.0 uses the old model for MPI and lacks a model
    for diffraction completely.}
  \label{fig:NSD2}
\end{figure*}

In Ref.~\cite{Khachatryan:2010us} a similar analysis was performed in
order to study the transverse momentum distributions of
non-single-diffractive events using the same corrections according
to the predictions by \Pythia{}. The new model shows a significant
improvement and seems to describe the data correctly except for the
ultra low $p_{\perp}<0.4 \,\mathrm{GeV}$ region (see
Fig.~\ref{fig:NSD2}).

\subsection{Underlying Event}

With the model for diffraction and the new model for soft interactions
at hand, \Herwig{}~7 for the first time attempts to give a satisfactory
description of minimum bias data. Before that we were limited to
diffraction reduced data samples.  The next important question is
whether the new model affects our previous description of the UE data
and possibly improve it. The UE is described as ``everything except the
hard scattering process'' and consists of contributions from the
initial- and final state radiation and hard and soft multiparticle
interactions. The measurements are made relative to a leading object
which is in this case the hardest charged track. In UE analyses three
regions of interest are usually considered. The threee regions are
defined according to their azimuthal angle with respect to the leading
track. The towards region, where $\phi <\pi/3$. The away region, where
$\phi >2\pi/3$ and the transverse region, where $\pi/3 < \phi < 2\pi/3$.
The towards and the away regions are usually the regions which are
dominated by the activity of the triggered hard scattering process. The
transverse region on the other hand contains little contribution from
the hard process and is therefore sensitive to interactions coming from
the UE.

\begin{figure*}[p]
  \begin{minipage}{0.3\textwidth} 
    \includegraphics[width=\textwidth]{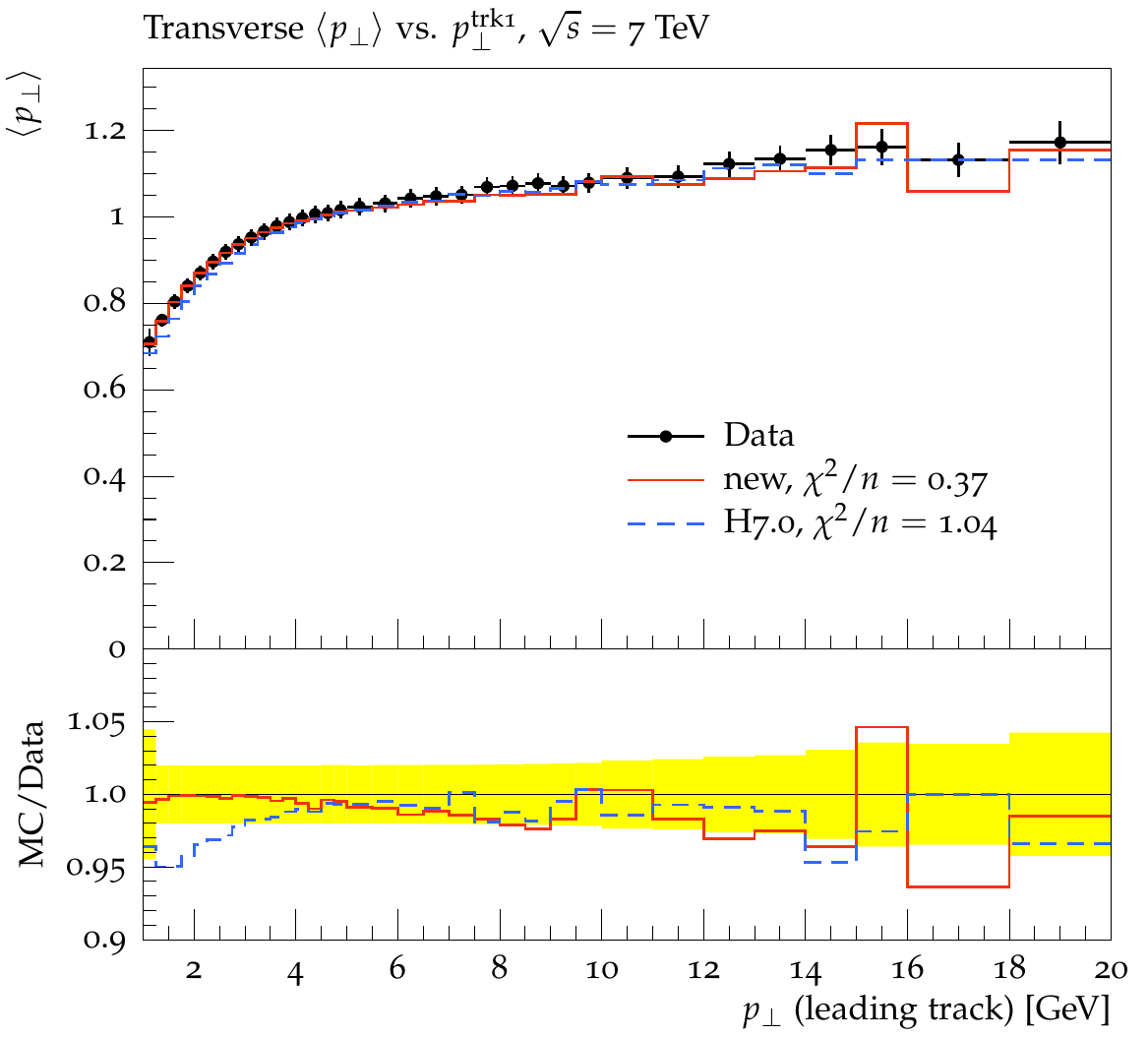}
  \end{minipage}
  \hfill
  \begin{minipage}{0.3\textwidth}
    \includegraphics[width=\textwidth]{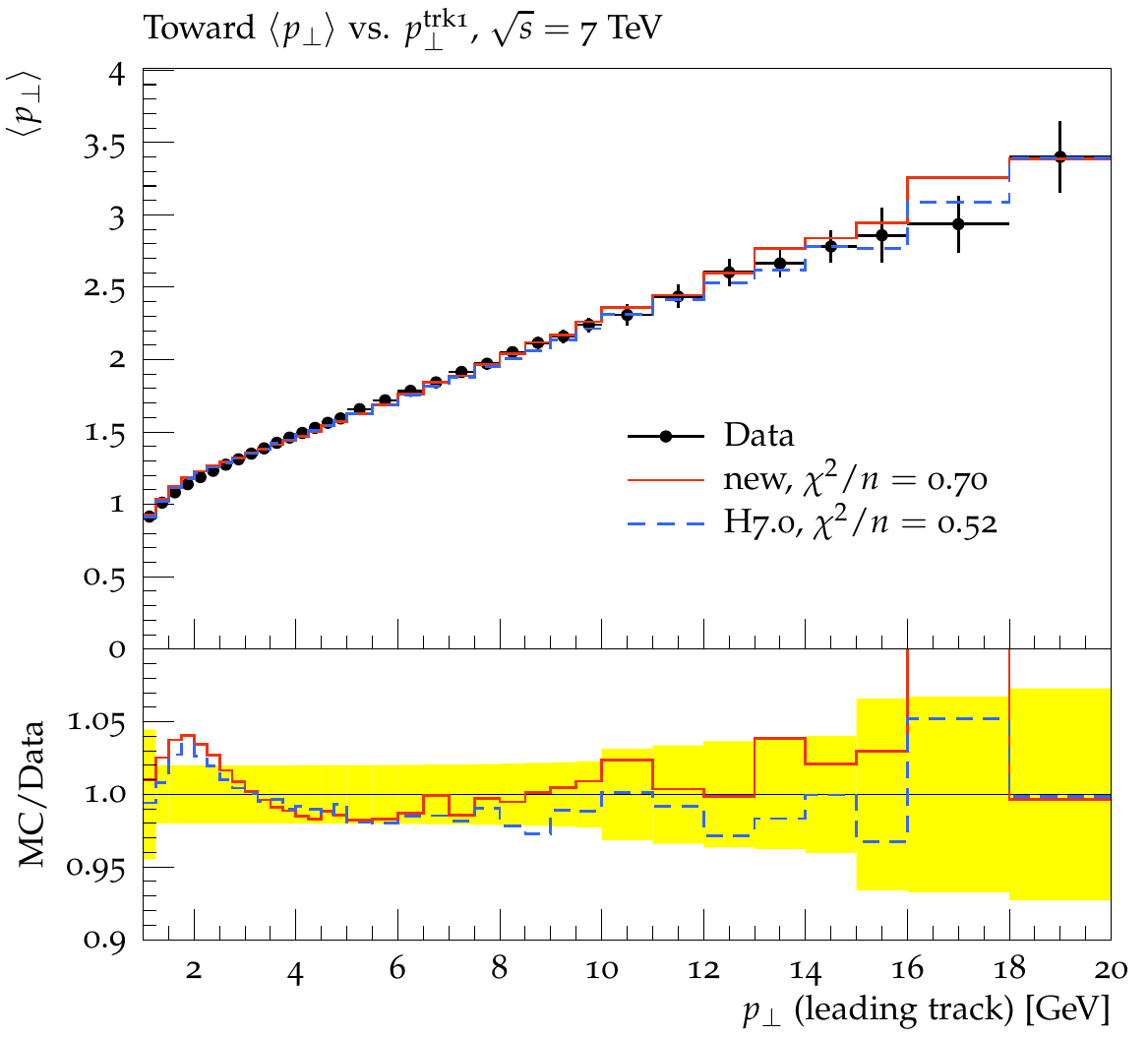}
  \end{minipage}
  \hfill
  \begin{minipage}{0.3\textwidth}
    \includegraphics[width=\textwidth]{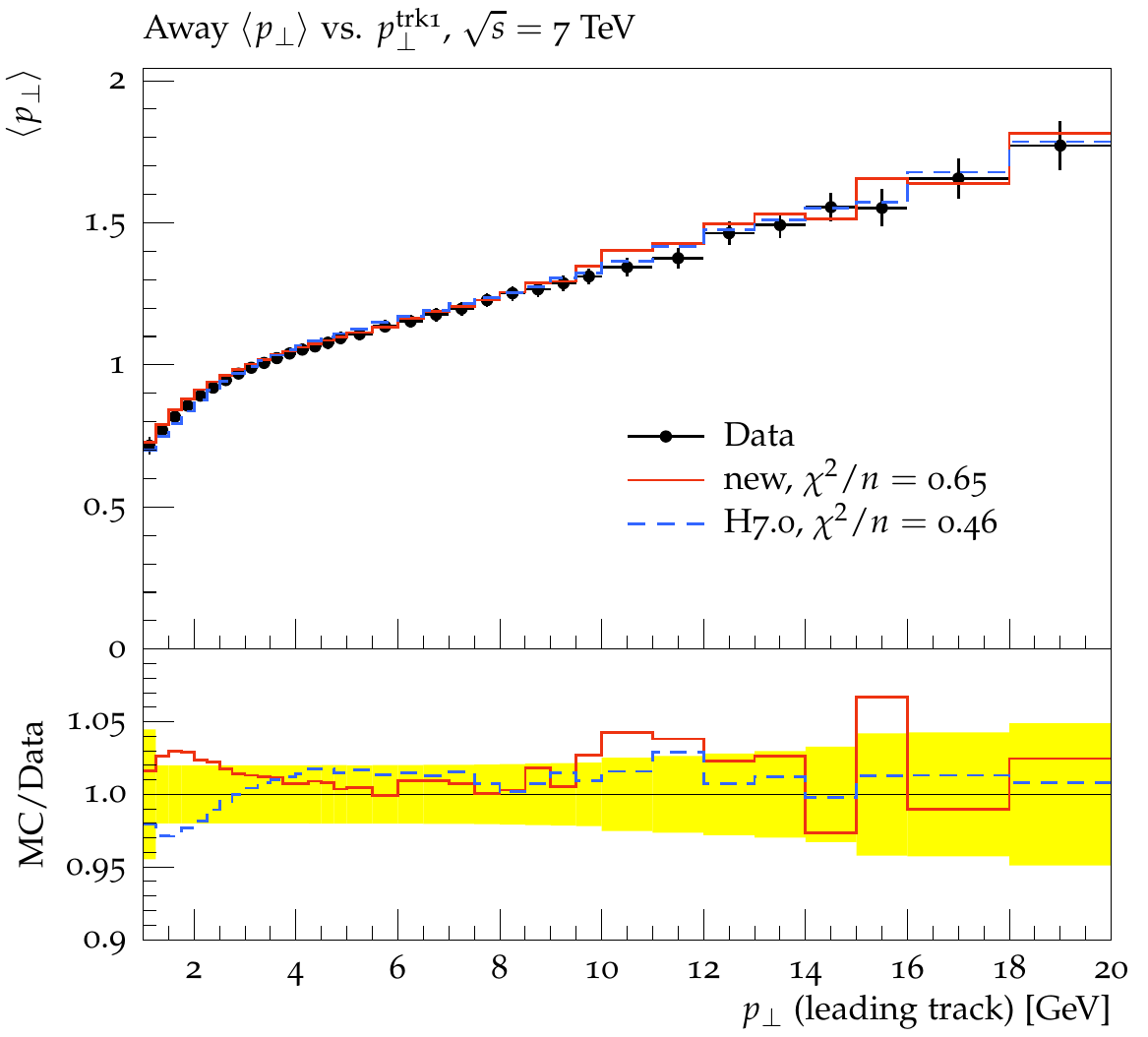}
  \end{minipage}
  \hfill
  \caption{Average transverse momentum $\langle p_{\perp}\rangle$ as a
    function of $p_{\perp}^{\mathrm{lead}}$ for the transverse, forward
    and away region. }
 \label{fig:UE1}
\end{figure*}

\begin{figure*}[p]
  \begin{minipage}{0.3\textwidth} 
    \includegraphics[width=\textwidth]{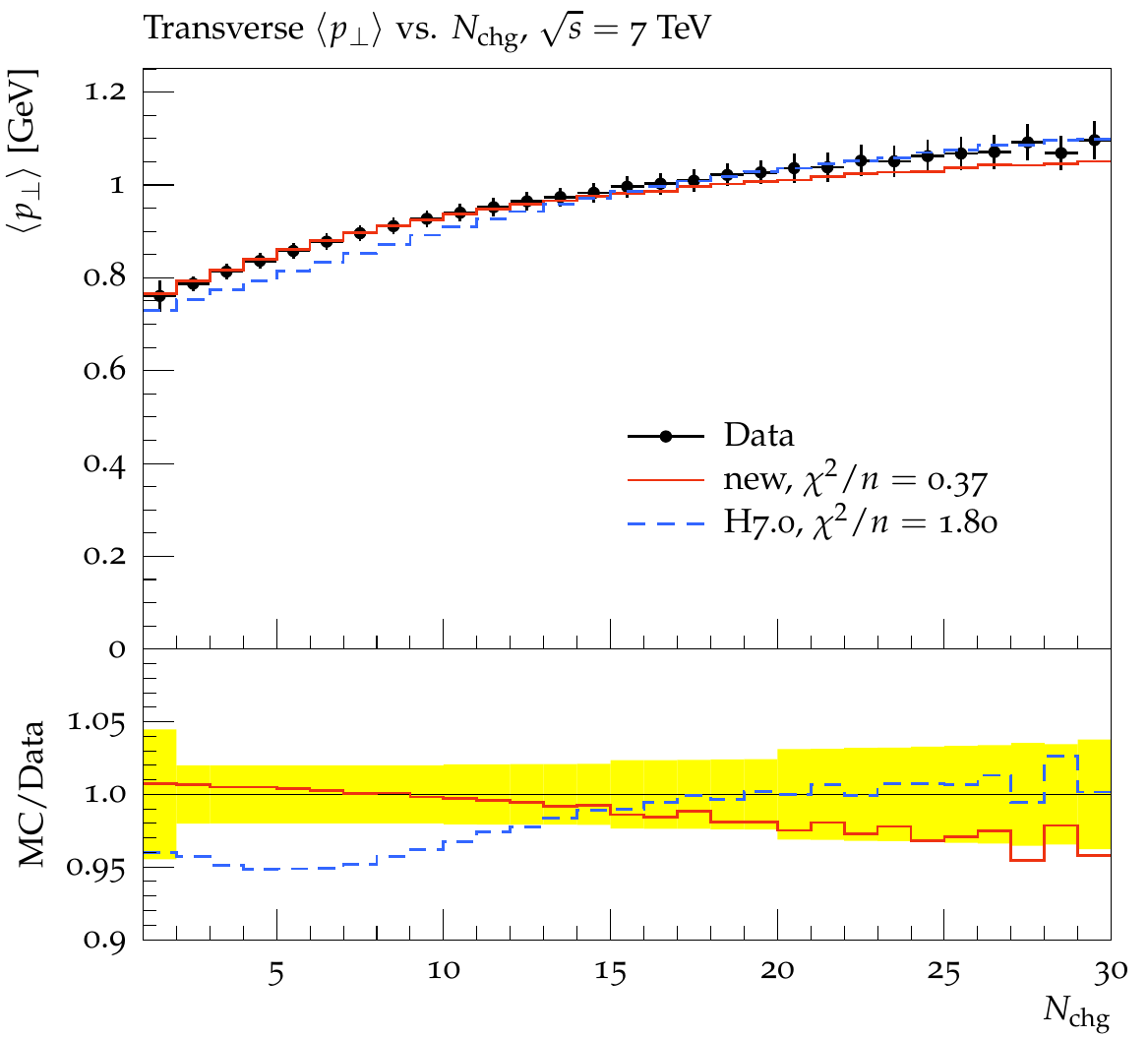}
  \end{minipage}
  \hfill
  \begin{minipage}{0.3\textwidth}
    \includegraphics[width=\textwidth]{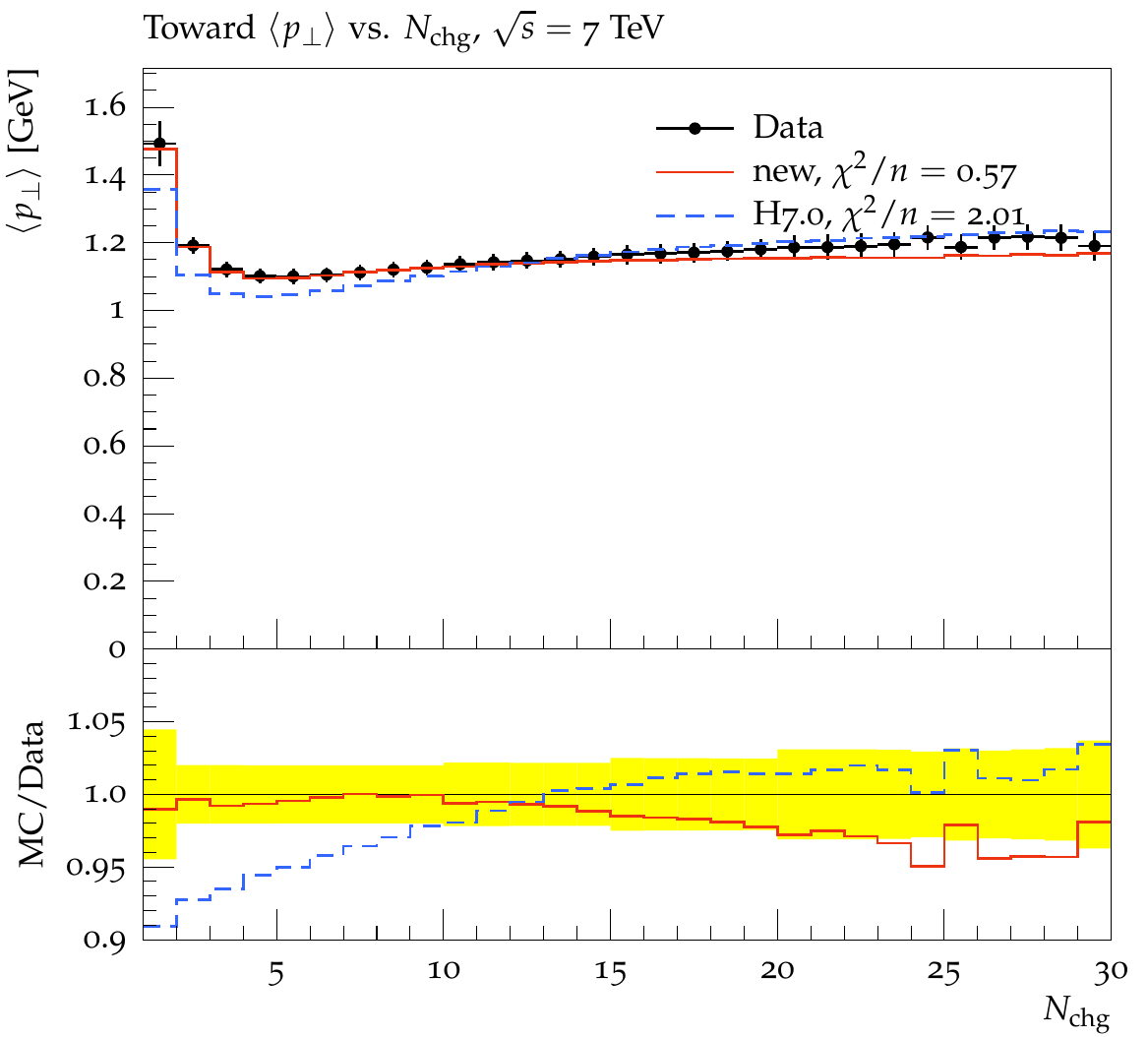}
  \end{minipage}
  \hfill
  \begin{minipage}{0.3\textwidth}
    \includegraphics[width=\textwidth]{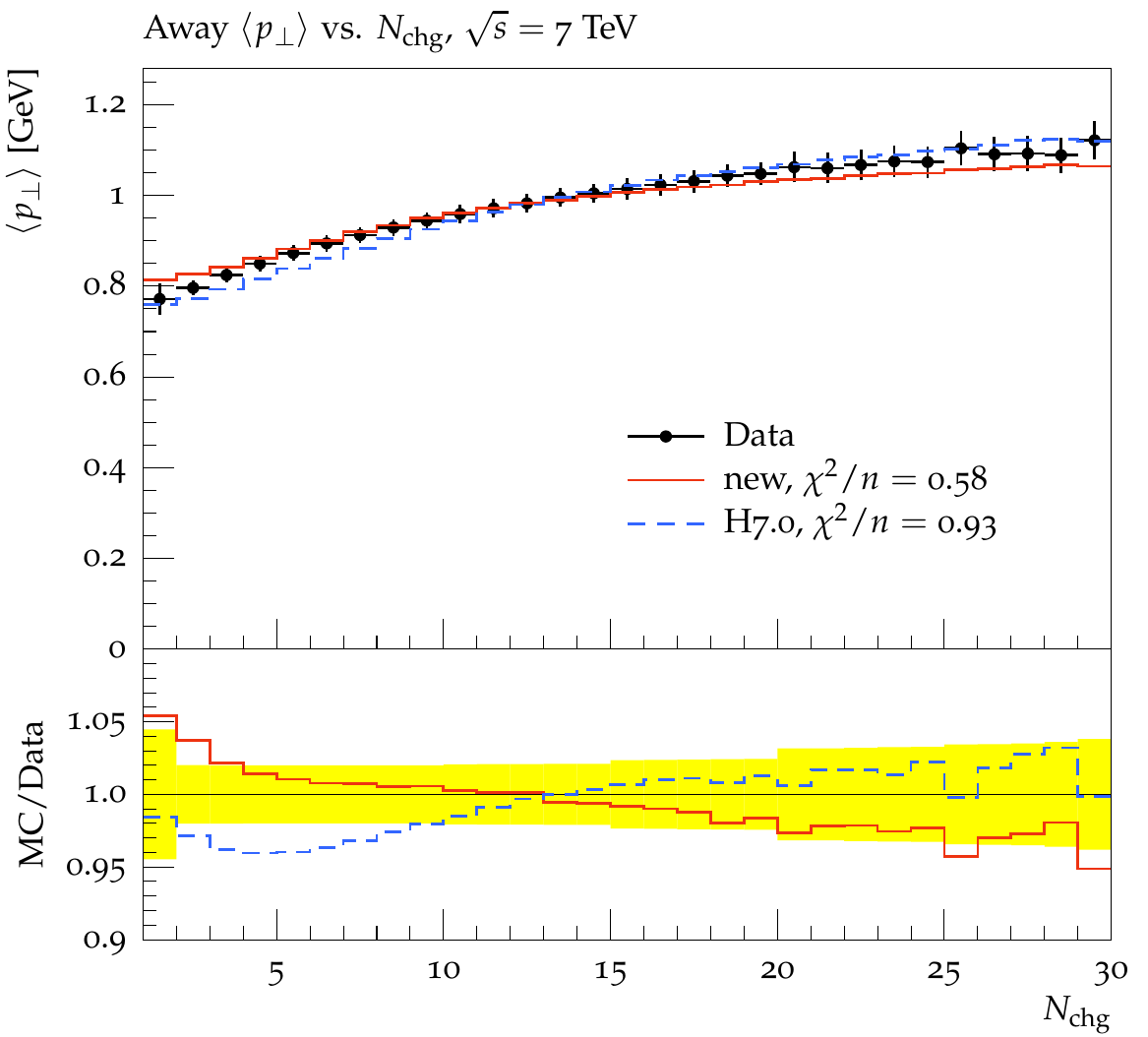}
  \end{minipage}
  \hfill
  \caption{Average transverse momentum $p_{\perp}$ over number of
    charged particles $N_{\rm ch}$ for the transverse, toward and away
    region. The data is compared with the new model and \Herwig{}~7. }
 \label{fig:UE2}
\end{figure*}

\begin{figure*}[p]
  \begin{minipage}{0.3\textwidth} 
    \includegraphics[width=\textwidth]{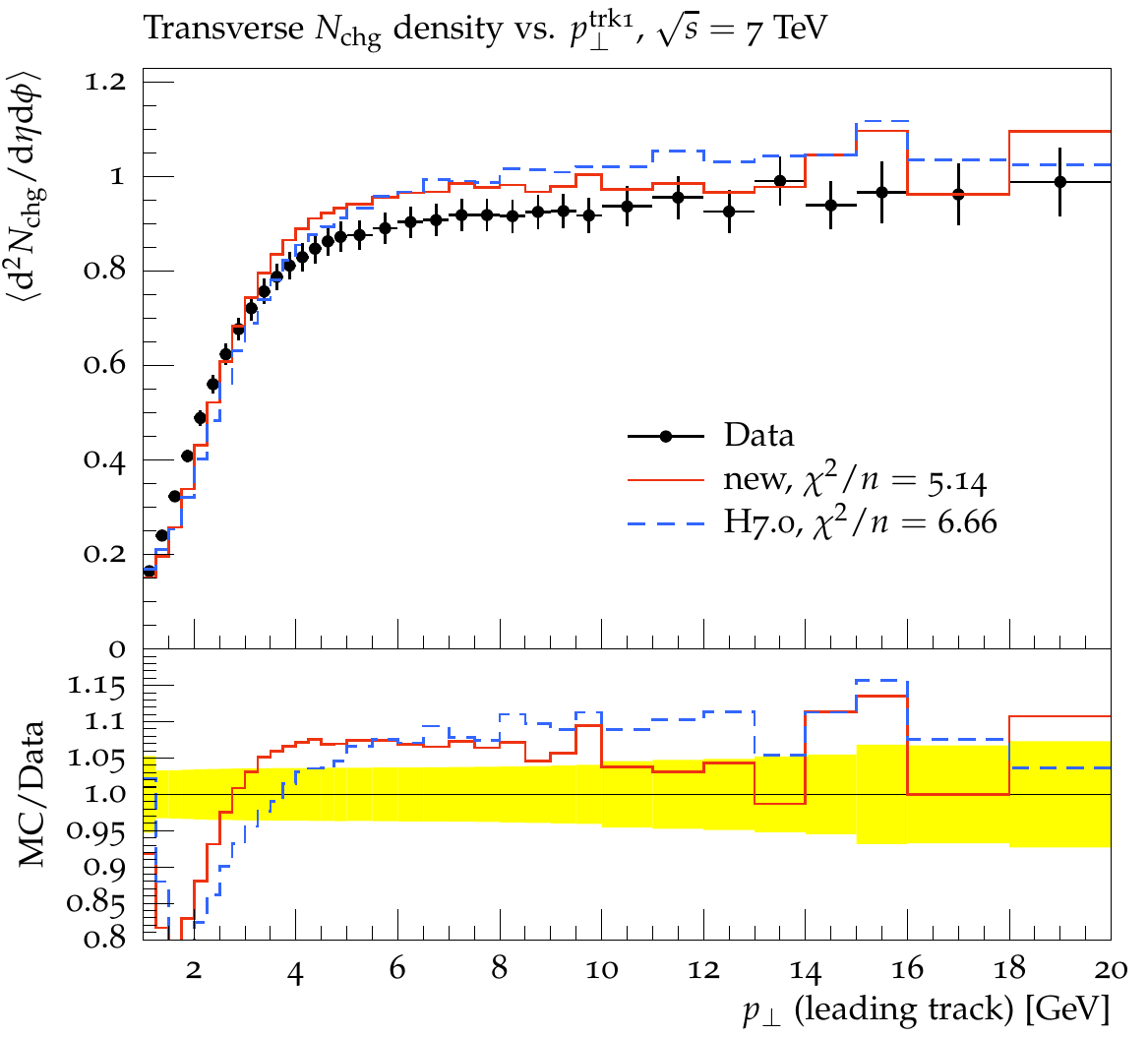}
  \end{minipage}
  \hfill
  \begin{minipage}{0.3\textwidth}
    \includegraphics[width=\textwidth]{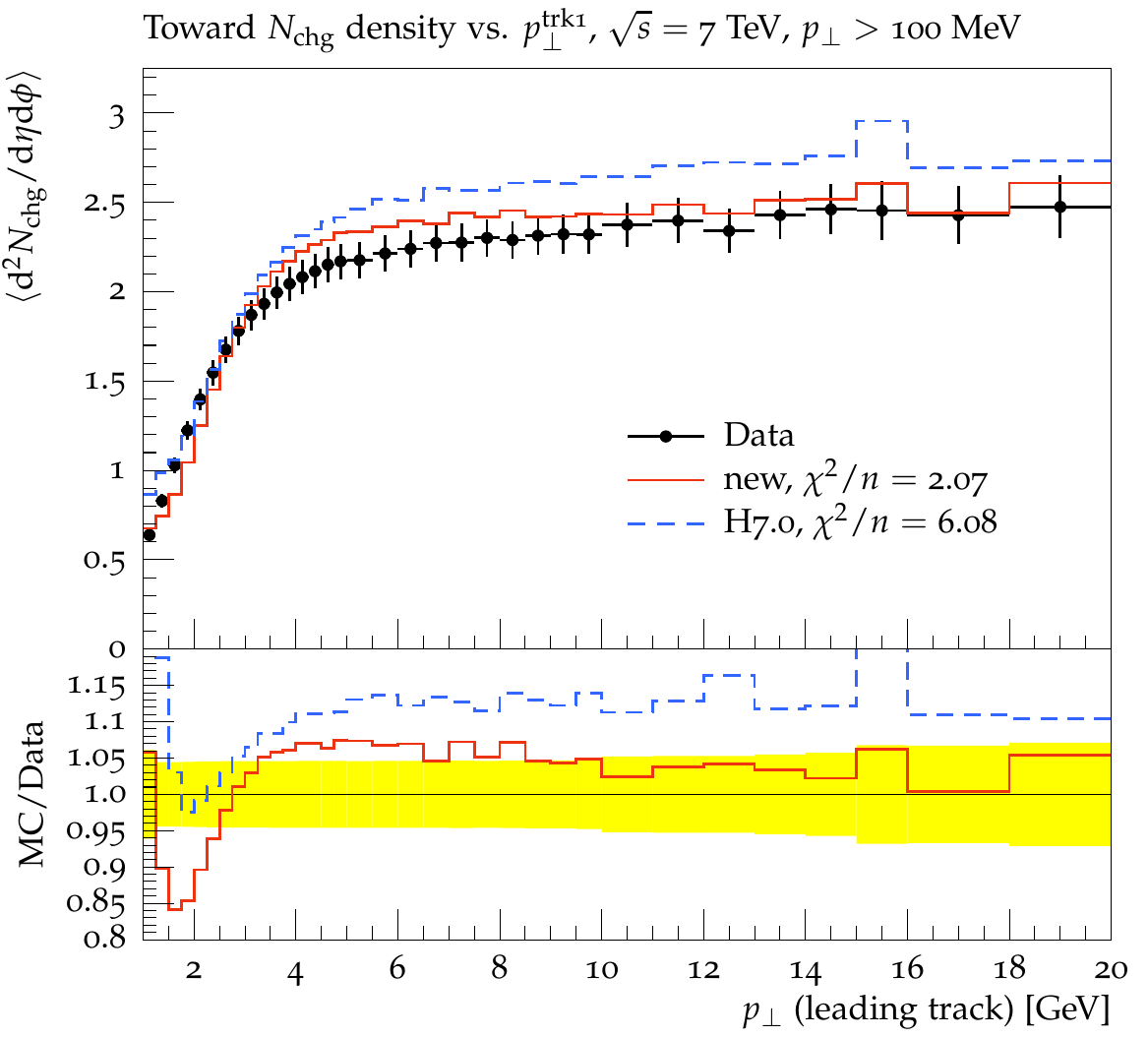}
  \end{minipage}
  \hfill
  \begin{minipage}{0.3\textwidth}
    \includegraphics[width=\textwidth]{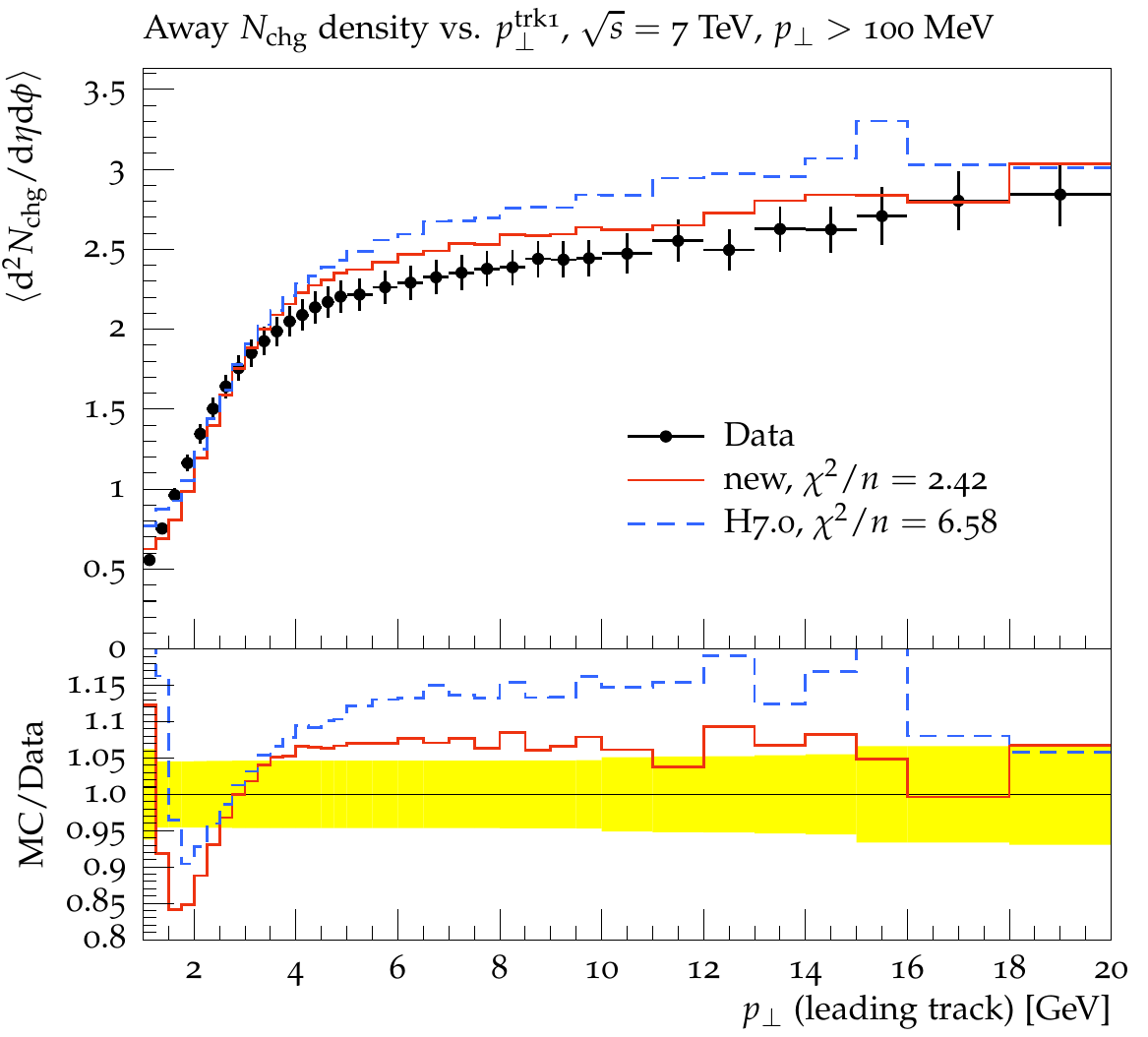}
  \end{minipage}
  \hfill
  \begin{minipage}{0.3\textwidth}
    \includegraphics[width=\textwidth]{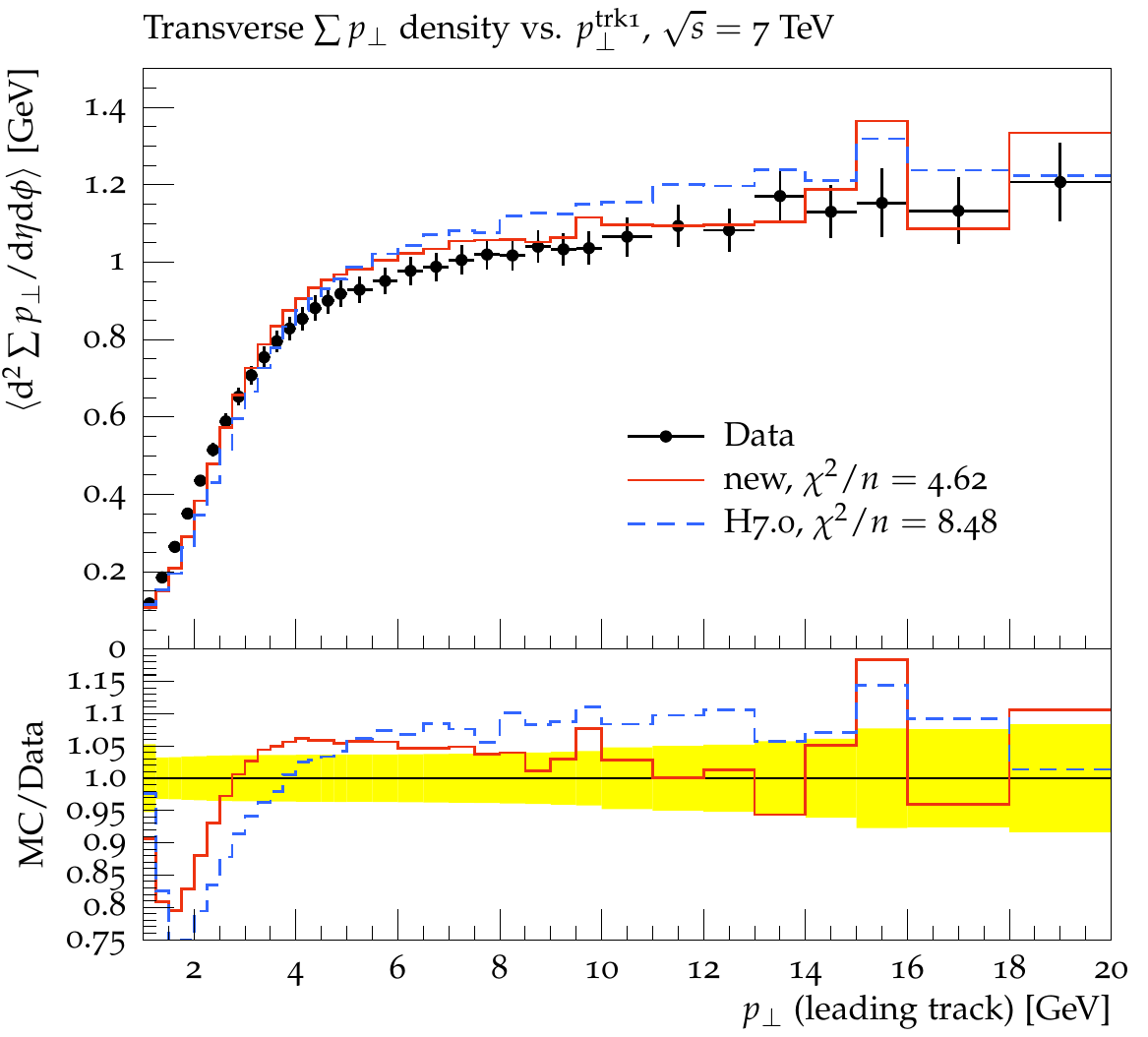}
  \end{minipage}
  \hfill
  \begin{minipage}{0.3\textwidth}
    \includegraphics[width=\textwidth]{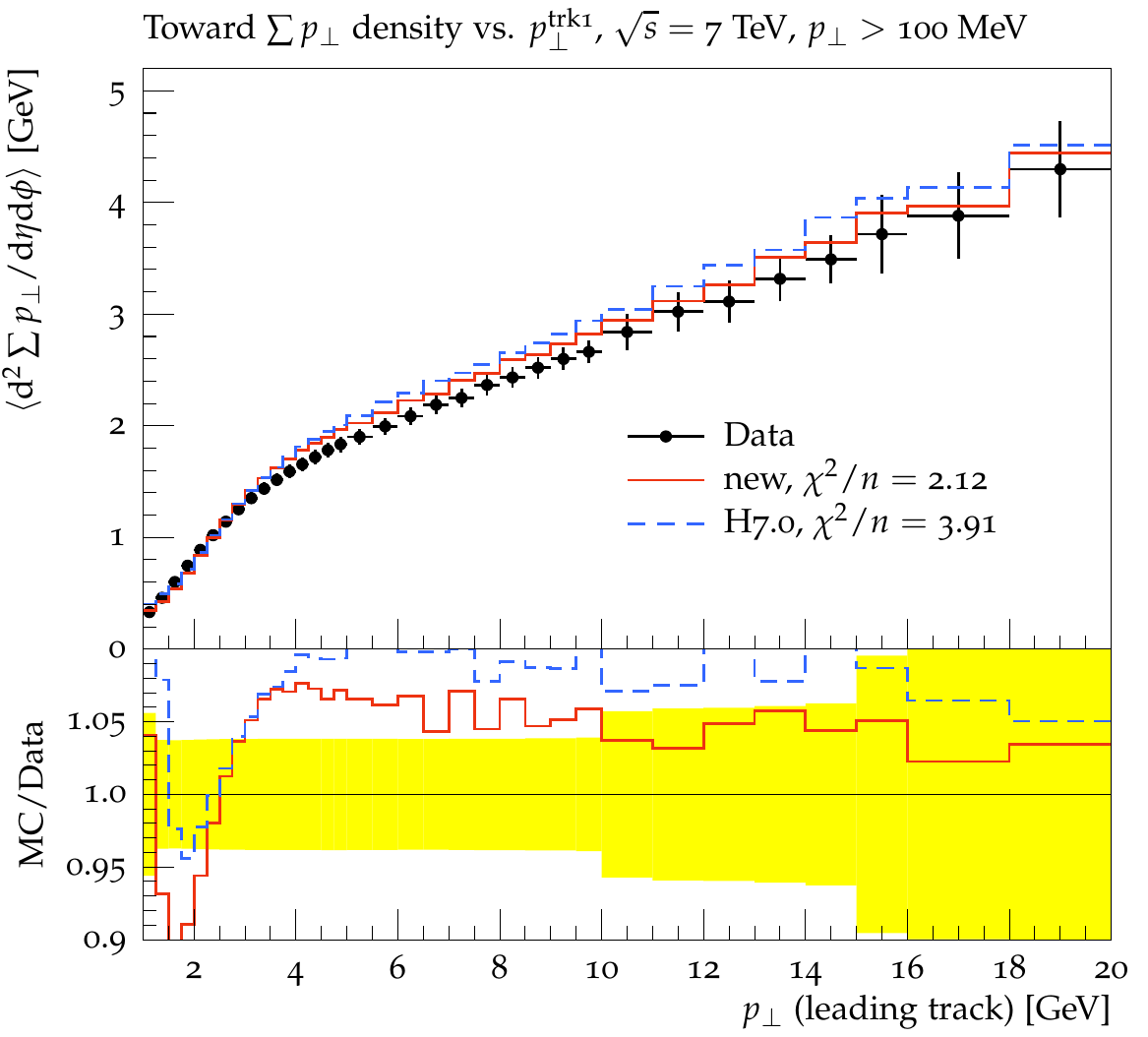}
  \end{minipage}
  \hfill
  \begin{minipage}{0.3\textwidth}
    \includegraphics[width=\textwidth]{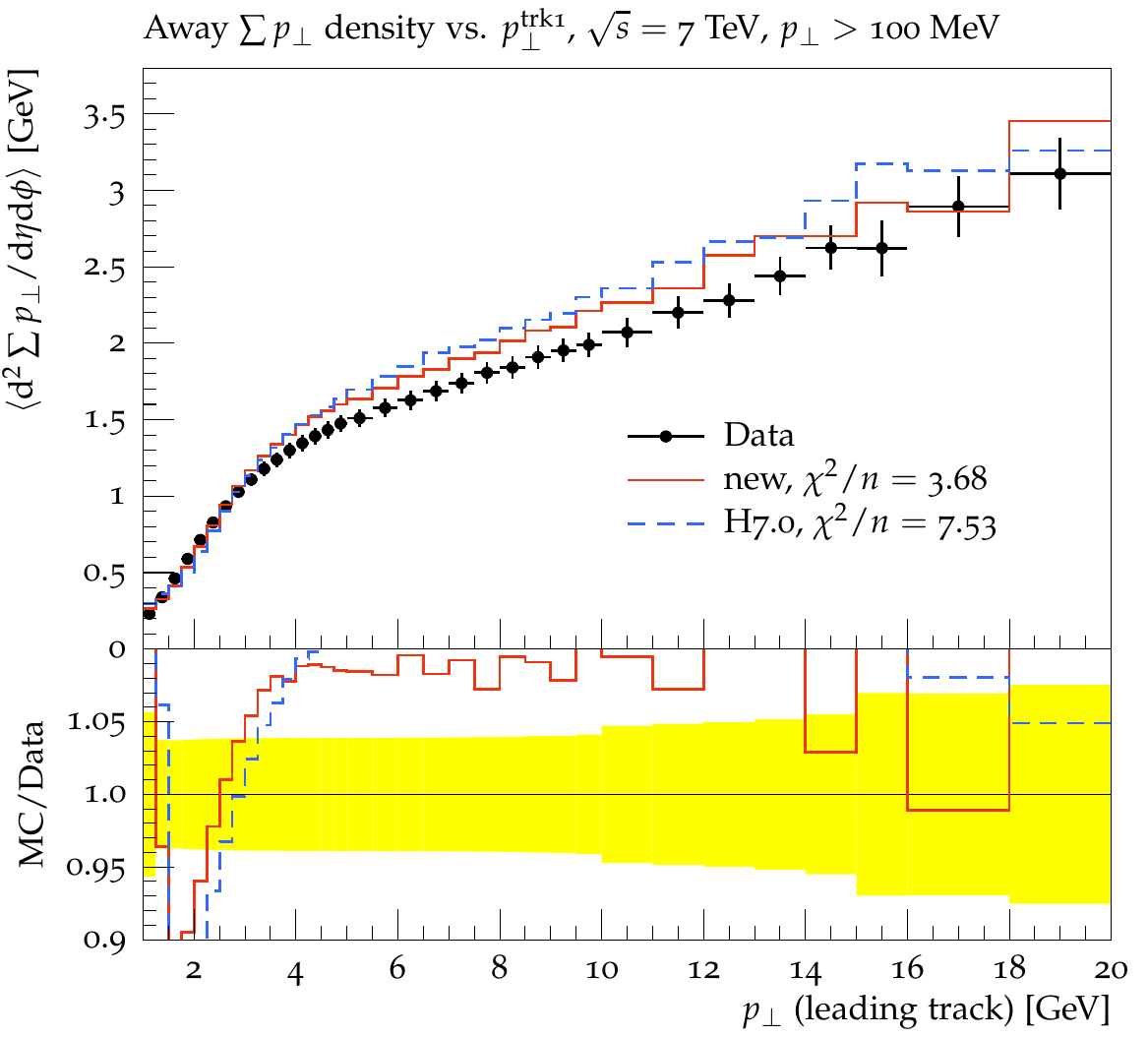}
  \end{minipage}
  \caption{Comparison of the default tune from \Herwig{}~7 with the new
    model to minimum-bias data from ATLAS \cite{Aad:2010ac} at
    $\sqrt{s} = 7 \, \mathrm{TeV}$.}
  \label{fig:UE3}
\end{figure*}

In Figs. \ref{fig:UE1} and \ref{fig:UE2} we show the
$\langle p_{\perp}\rangle$ distributions as a function of
$p_{\perp}^{\mathrm{lead}}$ and $N_{\mathrm{chg}}$. We see that in all
three regions, transverse, towards and away, the data is described
fairly well.  In Fig.~\ref{fig:UE3} we show more comparisons with UE
measurements from ATLAS.  This time we compare the number of charged
particles and the sum of transverse momenta in the three different
regions against data and our old model.  We find that the description
has improved for all observables. 

\subsection{Extrapolation to 13 TeV}
\begin{figure*}[p]
  \begin{minipage}{0.5\textwidth} 
    \includegraphics[width=1\textwidth]{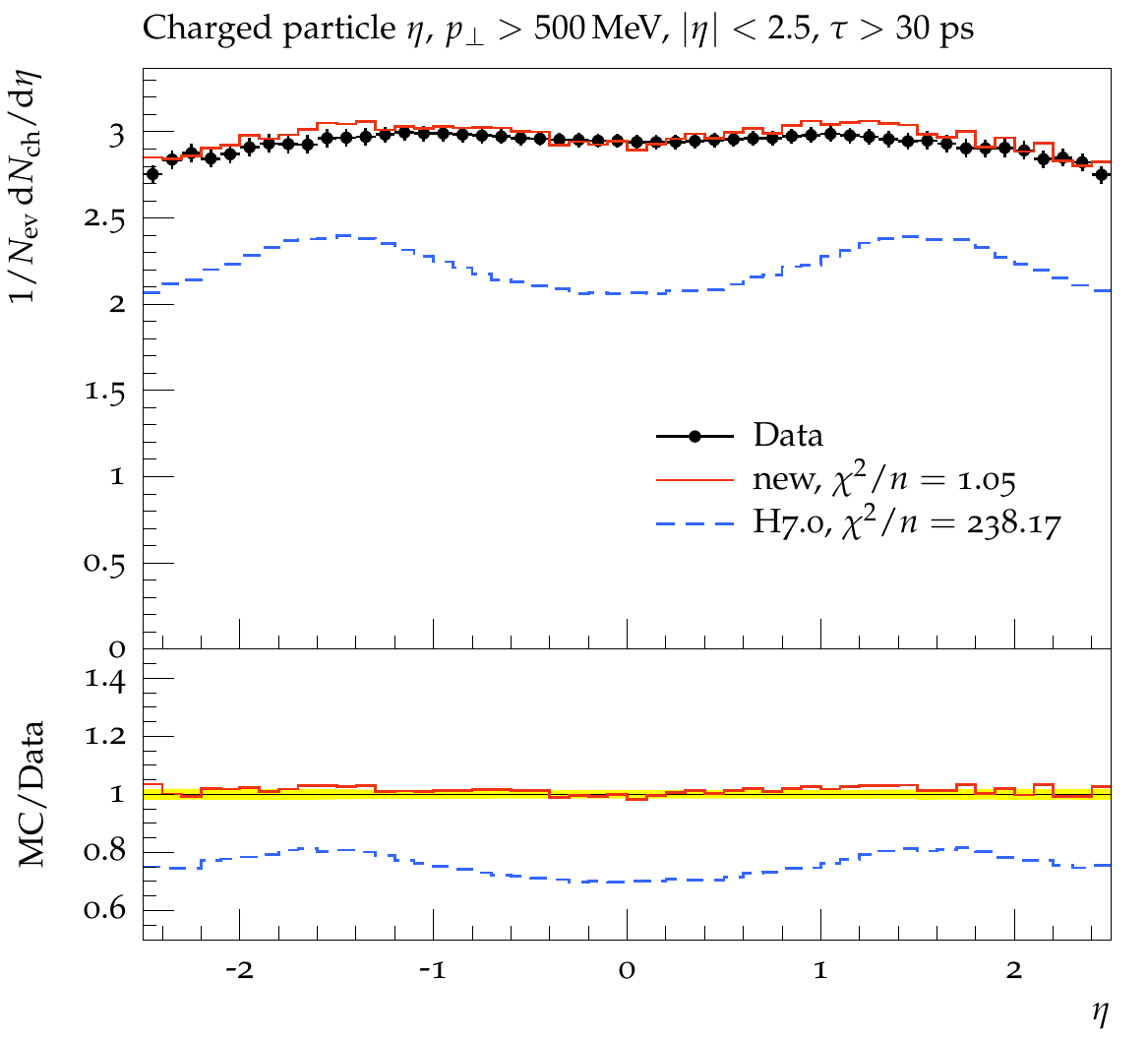}
  \end{minipage}
  \hfill
  \begin{minipage}{0.5\textwidth}
    
    \includegraphics[width=1\textwidth]{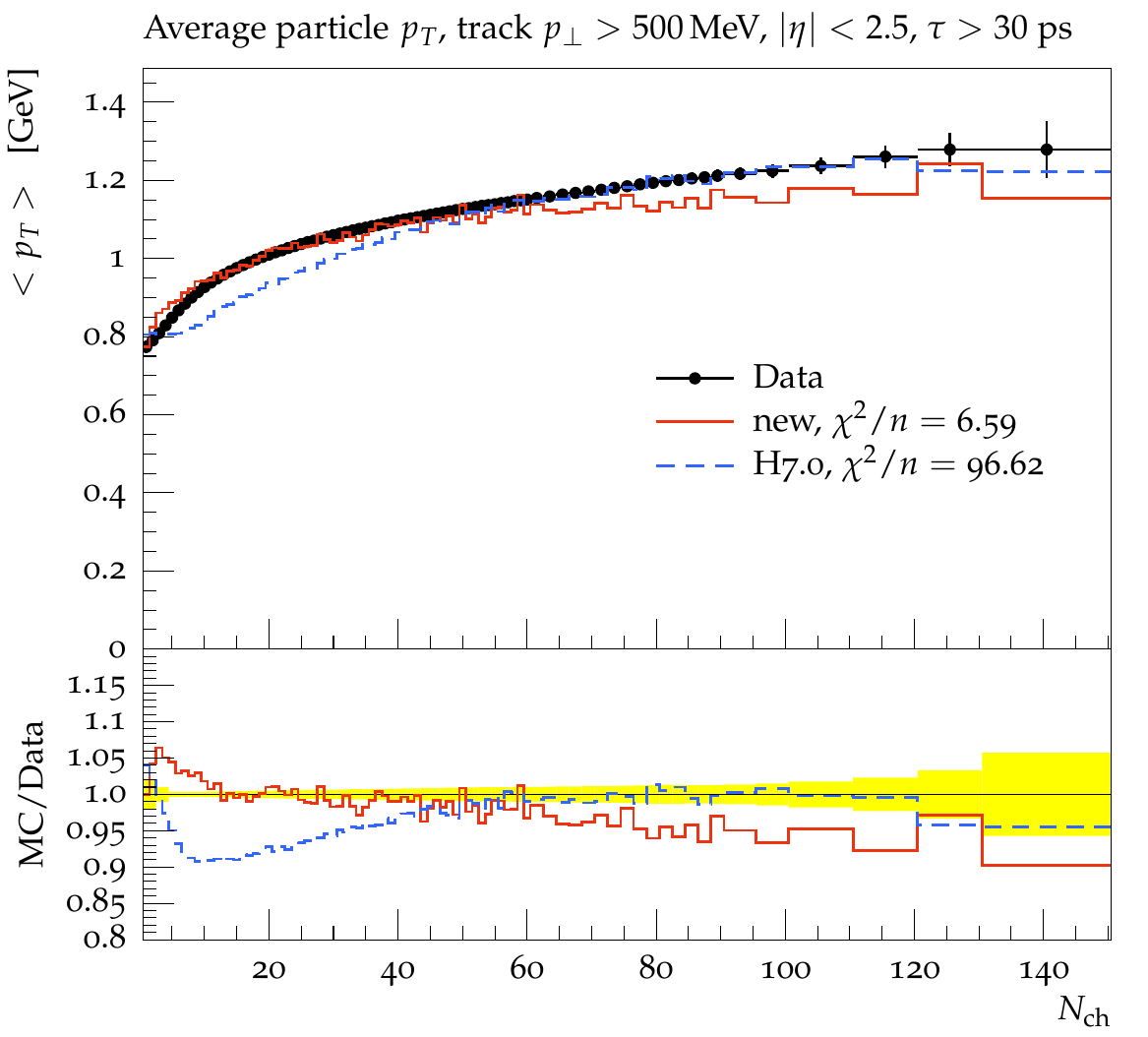}
  \end{minipage}
  \caption{Most inclusive $\eta$ distribution for
    $p_{\perp}>500 \, \mathrm{MeV}$ and average $p_{\perp}$ distribution
    for all particles with $p_{\perp}>500 \, \mathrm{MeV}$ measured by
    ATLAS \cite{Aad:2016mok} at $\sqrt{s} = 13 \, \mathrm{TeV}$. The
    runs for the new model were simulated with the tuned set of
    parameters for $7 \, \mathrm{TeV}$. H7.0 uses the old model for
    MPI.}
 \label{fig:13TeV}
\end{figure*}

With the energy update of the LHC to $13\, \mathrm{TeV}$ in 2015 new
sets of data are available. This data at the new energy frontier serves
as an excellent cross check for our new model. In order to test the
energy extrapolation we compare it to data provided by the ATLAS
collaboration \cite{Aad:2016mok} at $\sqrt{s} = 13 \, \mathrm{TeV}$.  We
used the same set of parameters as for $\sqrt{s} = 7\, \mathrm{TeV}$,
and did not tune the model parameters to any data taken at this energy,
the new model improves the description of the data compared to the old
model significantly as shown in Fig.~\ref{fig:13TeV}.

As a cross check, we also retuned the model at 13\,TeV.  As this
resulted in almost identical values for the parameters as for
$7 \, \mathrm{TeV}$, we have an excellent indication of a stable overall
energy scaling of our model.

\section{Summary and Outlook}
We have implemented a completely new model for soft physics in
\Herwig{}, which will become available with the next release of the
program.  A simple model for diffractive final states, based on the
cluster model is implemented in combination with a new model for
multiparticle production in soft interactions, based on multiperipheral
particle production.  We tuned the free parameters to data from
mini\-mum-bias measurements at 900\,MeV and 7\,TeV and obtained good
results in all observables for charged particles. Particularly the
rapidity gap observable, which has revealed a peculiar bump structure in
our previous model, is now well described. The quality of other
observables not considered in the tuning procedure is significantly
improved as well. We note that with these new models, \Herwig{}~7 is for
the first time able to describe the full range of minimum bias analyses
completely.  A stable extrapolation of our model to higher energies is
implied by a good description of 13\,TeV data albeit we did not tune to
data taken at this energy.

Remaining shortcomings of our model include the failure to describe the
average transverse momentum of all charged particles versus the event
multiplicity for very soft particles.  This hints at problematic
particle correlations via colour reconnections for very soft particles
or a problematic assignment of colour connections in the first place.
These shortcomings will be addressed in future work.  

Despite these remaining problems we regard this work as an important
first step forward in order to be able to describe collider phenomena
involving very soft particles for the first time.

\section{Acknowledgments}
We are grateful for helpful discussions on various aspects of this work
with several people, in particular Leif L\"{o}nnblad, Miroslav Myska,
Simon Pl\"{a}tzer, Agustin Sabio-Vera, Mike Seymour, Andrzej
Si\'{o}dmok.  FL thanks Y.~Dokshitzer and S.~Ostapchenko for interesting
discussions on this subject at various conferences.  This work was
supported in part by the European Union as part of the FP7 Marie Curie
Initial Training Network MCnetITN (PITN-GA-2012-315877).  Last but not
least we would like to thank the other members of the \Herwig{}
collaboration for continuous discussions and support.  


\begin{thebibliography}{10}

\bibitem{Bahr:2008pv}
{M. B\"ahr et al.},
\newblock Eur. Phys. J. {\bf C58}, 639 (2008), 0803.0883.

\bibitem{Sjostrand:2007gs}
T.~Sj{\"o}strand, S.~Mrenna, and P.~Skands,
\newblock Comput. Phys. Commun. {\bf 178}, 852 (2008), 0710.3820.

\bibitem{Gleisberg:2008ta}
T.~Gleisberg {\em et~al.},
\newblock JHEP {\bf 0902}, 007 (2009), 0811.4622.

\bibitem{Fischer:2016zzs}
N.~Fischer and T.~Sjöstrand,
\newblock (2016), 1610.09818.

\bibitem{Rasmussen:2015qgr}
C.~O. Rasmussen and T.~Sjöstrand,
\newblock JHEP {\bf 02}, 142 (2016), 1512.05525.

\bibitem{Christiansen:2015yca}
J.~R. Christiansen and T.~Sjöstrand,
\newblock Eur. Phys. J. {\bf C75}, 441 (2015), 1506.09085.

\bibitem{Bierlich:2015rha}
C.~Bierlich and J.~R. Christiansen,
\newblock Phys. Rev. {\bf D92}, 094010 (2015), 1507.02091.

\bibitem{Hoche:2007hg}
S.~H{\"o}che, F.~Krauss, and T.~Teubner,
\newblock Eur. Phys. J. {\bf C58}, 17 (2008), 0705.4577.

\bibitem{Sjostrand:1987su}
T.~Sj{\"o}strand and M.~van Zijl,
\newblock Phys.Rev. {\bf D36}, 2019 (1987).

\bibitem{Butterworth:1996zw}
J.~Butterworth, J.~R. Forshaw, and M.~Seymour,
\newblock Z.Phys. {\bf C72}, 637 (1996), hep-ph/9601371.

\bibitem{Bahr:2008wk}
M.~B{\"a}hr, J.~M. Butterworth, and M.~H. Seymour,
\newblock JHEP {\bf 0901}, 065 (2009), 0806.2949.

\bibitem{Bahr:2008dy}
M.~B{\"a}hr, S.~Gieseke, and M.~H. Seymour,
\newblock JHEP {\bf 0807}, 076 (2008), 0803.3633.

\bibitem{Borozan:2002fk}
I.~Borozan and M.~Seymour,
\newblock JHEP {\bf 0209}, 015 (2002), hep-ph/0207283.

\bibitem{Bahr:2009ek}
M.~B{\"a}hr, J.~M. Butterworth, S.~Gieseke, and M.~H. Seymour,
\newblock (2009), 0905.4671.

\bibitem{Aad:2012pw}
ATLAS, G.~Aad {\em et~al.},
\newblock Eur. Phys. J. {\bf C72}, 1926 (2012), 1201.2808.

\bibitem{Gieseke:2016pbi}
S.~Gieseke, F.~Loshaj, and M.~Myska,
\newblock {Towards Diffraction in Herwig},
\newblock in {\em {Proceedings, 7th International Workshop on Multiple Partonic
  Interactions at the LHC (MPI@LHC 2015): Miramare, Trieste, Italy, November
  23-27, 2015}}, pp. 53--57, 2016, 1602.04690.

\bibitem{Ciafaloni:1987ur}
M.~Ciafaloni,
\newblock Nucl. Phys. {\bf B296}, 49 (1988).

\bibitem{Catani:1989sg}
S.~Catani, F.~Fiorani, and G.~Marchesini,
\newblock Nucl. Phys. {\bf B336}, 18 (1990).

\bibitem{Catani:1989yc}
S.~Catani, F.~Fiorani, and G.~Marchesini,
\newblock Phys. Lett. {\bf B234}, 339 (1990).

\bibitem{Marchesini:1994wr}
G.~Marchesini,
\newblock Nucl. Phys. {\bf B445}, 49 (1995), hep-ph/9412327.

\bibitem{Jung:2001hx}
H.~Jung,
\newblock Comput. Phys. Commun. {\bf 143}, 100 (2002), hep-ph/0109102.

\bibitem{Balitsky:1978ic}
I.~I. Balitsky and L.~N. Lipatov,
\newblock Sov. J. Nucl. Phys. {\bf 28}, 822 (1978).

\bibitem{Mueller:1970fa}
A.~H. Mueller,
\newblock Phys. Rev. {\bf D2}, 2963 (1970).

\bibitem{Barone:2002cv}
V.~Barone and E.~Predazzi,
\newblock {\em {High-Energy Particle Diffraction}}, Texts and Monographs in
  Physics Vol. v.565 (Springer-Verlag, Berlin Heidelberg, 2002).

\bibitem{Abelev:2012sea}
ALICE, B.~Abelev {\em et~al.},
\newblock Eur. Phys. J. {\bf C73}, 2456 (2013), 1208.4968.

\bibitem{Amati:1962nv}
D.~Amati, A.~Stanghellini, and S.~Fubini,
\newblock Nuovo Cim. {\bf 26}, 896 (1962).

\bibitem{Baker:1976cv}
M.~Baker and K.~A. Ter-Martirosian,
\newblock Phys. Rept. {\bf 28}, 1 (1976).

\bibitem{Buckley:2009bj}
A.~Buckley, H.~Hoeth, H.~Lacker, H.~Schulz, and J.~E. von Seggern,
\newblock Eur. Phys. J. {\bf C65}, 331 (2010), 0907.2973.

\bibitem{Gieseke:2012ft}
S.~Gieseke, C.~R{\"o}hr, and A.~Siodmok,
\newblock Eur.Phys.J. {\bf C72}, 2225 (2012), 1206.0041.

\bibitem{Aad:2010ac}
ATLAS Collaboration, G.~Aad {\em et~al.},
\newblock New J.Phys. {\bf 13}, 053033 (2011), 1012.5104.

\bibitem{Khachatryan:2015gka}
CMS, V.~Khachatryan {\em et~al.},
\newblock Phys. Rev. {\bf D92}, 012003 (2015), 1503.08689.

\bibitem{Khachatryan:2010nk}
CMS Collaboration, V.~Khachatryan {\em et~al.},
\newblock JHEP {\bf 1101}, 079 (2011), 1011.5531.

\bibitem{Sjostrand:2006za}
T.~Sj{\"o}strand, S.~Mrenna, and P.~Skands,
\newblock JHEP {\bf 0605}, 026 (2006), hep-ph/0603175.

\bibitem{Khachatryan:2010us}
CMS Collaboration, V.~Khachatryan {\em et~al.},
\newblock Phys.Rev.Lett. {\bf 105}, 022002 (2010), 1005.3299.

\bibitem{Aad:2016mok}
ATLAS, G.~Aad {\em et~al.},
\newblock Phys. Lett. {\bf B758}, 67 (2016), 1602.01633.

\end{thebibliography}

\end{document}